\documentclass[twocolumn,english,reprint,superscriptaddress,prl]{revtex4-1}

\usepackage[T1]{fontenc}
\usepackage[latin9]{inputenc}
\usepackage{color}
\usepackage{bm}
\usepackage{amstext}
\usepackage{amssymb}
\usepackage{graphicx}
\usepackage{booktabs}
\usepackage{tabularx}
\usepackage{amsmath}
 \usepackage{algpseudocode}

\makeatletter

\usepackage{soul}
\newcommand{\pt}{p_{\boldsymbol{\lambda}}}
\newcommand{\pc}{\tilde{p}_{\boldsymbol{\lambda}}}

\newcommand{\ket}[1]{| #1 \rangle}
\newcommand{\bra}[1]{\langle #1 |}
\newcommand{\rb}{$^{87}$Rb }
\newcommand{\pfd}{P_{\textnormal{FD}}}
\newcommand{\pgt}{P_{\textnormal{GT}}}

\usepackage{babel}

\usepackage{wasysym}

\newcommand{\bsig}{\bm{\sigma}}
\newcommand{\btau}{\bm{\tau}}
\newcommand{\blda}{\bm{\lambda}}

\newcommand{\bh}{\bm{h}}
\newcommand{\bb}{\bm{b}}
\newcommand{\tr}{\textnormal{Tr}}

\newcommand{\waterloo}{Department of Physics and Astronomy, University of Waterloo, Ontario N2L 3G1, Canada}
\newcommand{\caltech}{Division of Physics, Mathematics and Astronomy, California Institute of Technology, Pasadena, CA 91125, USA}
\newcommand{\flatiron}{Center for Computational Quantum Physics, Flatiron Institute, New York, New York 10010, USA}
\newcommand{\perimeter}{Perimeter Institute for Theoretical Physics, Waterloo, Ontario N2L 2Y5, Canada}
\newcommand{\harvard}{Department of Physics, Harvard University, Cambridge, MA 02138, USA}
\newcommand{\mitrle}{Department of Physics and Research Laboratory of Electronics, Massachusetts Institute of Technology, Cambridge, MA 02139, USA}
\newcommand{\chicago}{Institute for Molecular Engineering, University of Chicago, Chicago, IL 60637, USA}

\makeatother

\begin{document}
\author{Giacomo Torlai}
\thanks{These authors contributed equally.}
\affiliation{\flatiron}
\affiliation{\waterloo}
\affiliation{\perimeter}
\author{Brian Timar}
\thanks{These authors contributed equally.}
\affiliation{\caltech}
\author{Evert P.L. van Nieuwenburg}
\affiliation{\caltech}
\author{Harry Levine}
\affiliation{\harvard}
\author{Ahmed Omran}
\affiliation{\harvard}
\author{Alexander Keesling}
\affiliation{\harvard}
\author{Hannes Bernien}
\affiliation{\chicago}
\author{Markus Greiner}
\affiliation{\harvard}
\author{Vladan Vuleti\'{c}}
\affiliation{\mitrle}
\author{Mikhail D. Lukin}
\affiliation{\harvard}
\author{Roger G. Melko}
\affiliation{\waterloo}
\affiliation{\perimeter}
\author{Manuel Endres}
\affiliation{\caltech}

\title{Integrating Neural Networks with a Quantum Simulator for State Reconstruction}

\begin{abstract}

We demonstrate quantum many-body state reconstruction from experimental data generated by a programmable quantum simulator, by means of a neural network model incorporating known experimental errors. Specifically, we extract restricted Boltzmann machine (RBM) wavefunctions from data produced by a Rydberg quantum simulator with eight and nine atoms in a single measurement basis, and apply a novel regularization technique to mitigate the effects of measurement errors in the training data. Reconstructions of modest complexity are able to capture one- and two-body observables not accessible to experimentalists, as well as more sophisticated observables such as the  R\'enyi mutual information. Our results open the door to integration of machine learning architectures with intermediate-scale quantum hardware.

\end{abstract}
\maketitle
\selectlanguage{english}
\setlength{\parskip}{2pt}

Quantum state tomography~\cite{Banaszek2013} is an important tool for reconstructing generic quantum states, but traditional techniques require a number of measurements scaling exponentially in the system size~\cite{Haffner2005}. In certain cases, methods that exploit particular entanglement or symmetry properties~\cite{Cramer2010, Lee2015, Riofrio2017, Lanyon2017, Toth2010} allow for more efficient tomography of states prepared in experiments. However, such approaches still involve explicit reconstruction of local density operators \cite{Jezek2003,Cramer2010}, incurring a significant computational overhead in the estimation of nontrivial observables from experimental data -- especially in the presence of measurement errors introduced by realistic experimental hardware. In order to facilitate the characterization of near-term quantum hardware~\cite{Preskill2018}, a state reconstruction method which can efficiently extract physical quantities of interest directly from noisy experimental datasets is highly desirable.

Neural network-based machine learning has recently emerged as a powerful technique for learning compact representations of high-dimensional data~\cite{Hinton2006AE, Graves2013, LeCun2015}. In experimental quantum science, these tools have already been applied profitably to the classification of experimental snapshots~\cite{Rem2018, Bohrdt2018} and qubit readout~\cite{Seif2018}. The same data-driven approach can be applied to tomographic tasks. Recent theoretical work has demonstrated that a generative model called a restricted Boltzmann machine (RBM) is capable of accurate reconstruction of quantum states and observables directly from synthetic datasets generated by numerical algorithms~\cite{rbm_nisq}.

\begin{figure}[t]
\noindent \centering \includegraphics[width=\columnwidth]{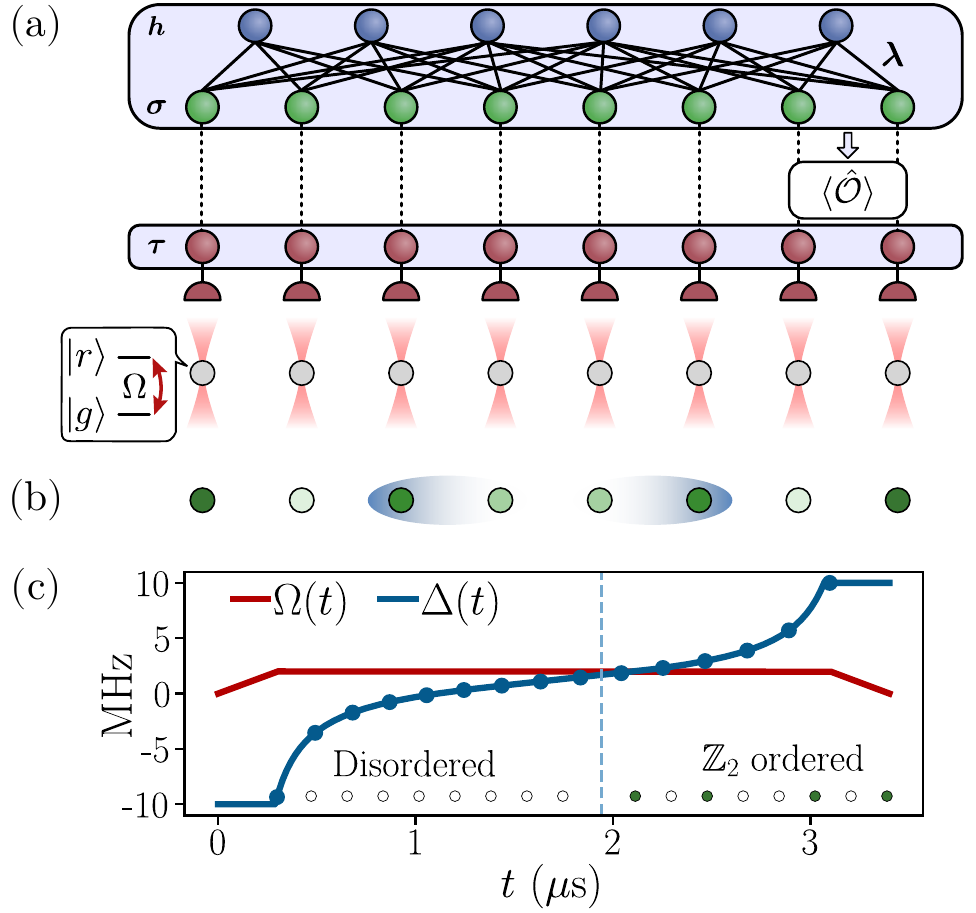}
\caption{{Experiment and reconstruction.} 
(a) Model of the reconstruction process. Individual \rb atoms (grey circles) are trapped in an array of optical tweezers and coupled to a Rydberg state with Rabi frequency $\Omega$. Site-resolved fluorescence imaging provides imperfect measurement in the $\hat \sigma^z$ basis. Our neural network model describes the true quantum state as an RBM (blue and green neurons), while the binary data $\btau$ accessible to the experimentalist are included as an auxiliary `noise' layer (red neurons). By training on this data, the network learns parameters $\blda$ describing the experimental quantum state, which are subsequently used to compute observables $\langle \hat{ \mathcal{ O} }\rangle$.
(b) Representation of the ordered state at the end of the adiabatic sweep -- see Eq~(\ref{eq:PsiGS}). Darker circles represent a higher probability of Rydberg excitation, and the shading indicates quantum fluctuations localized at
bonds (3,4) and (5,6). 
(c) The effective laser detuning $\Delta$ and Rabi frequency $\Omega$ as a function of sweep time $t$. Circular markers indicate the times at which the sweep was halted to collect data.
Vertical line: approximate transition to ordering in the finite system. The nearest-neighbor interaction is $V_{nn}=30$ MHz, the final detuning is 10 MHz, and the peak Rabi frequency is 2 MHz; the total sweep time is $T_{\textnormal{ev}} = 3.4\mu$s.
}
\label{Fig::experiment}
\end{figure}

In this Letter, we present a proof-of-principle demonstration of neural network quantum state reconstruction from experimental data. Our experimental system consists of a one-dimensional array of strongly interacting Rydberg atoms~\cite{Bernien2017, Endres2016}. Leveraging the high purity and approximate positivity of the experimental state, we train RBMs using single measurement basis data consisting of bit-strings obtained via repeated, simultaneous single-shot readout of the ground and Rydberg populations of all atoms. The RBMs learn a higher-fidelity and more efficient representation of the underlying bit-string probability distributions than standard inference from the limited size training dataset. This approach also enables us to implement an efficient procedure for denoising the full probability distribution from bit-flip-type measurement errors, by incorporating a dedicated ``noise layer'' in the network architecture. We test the validity of our approach by comparing predictions of the trained RBMs with numerical results for observables that are off-diagonal in the measurement basis, including the quantum mutual information. These results demonstrate the utility of RBMs in reconstructing approximately pure, positive states from experimental data, and pave the way to further integration of neural network models with quantum hardware.

\paragraph{Experimental system.} 
Our experimental approach~\cite{Bernien2017, Endres2016} involves a programmable Rydberg atom quantum simulator, a flexible neutral-atom system for realizing Ising-type quantum spin models~\cite{Schauss2015, Labuhn2016,Zeiher2016, Zeiher2017, Bernien2017,Guardado-Sanchez2018, Barredo2018}. In the present experiments 
(Fig.~\ref{Fig::experiment}a), a one-dimensional array of $N$ trapped Rubidium atoms is prepared; $N=8$ atoms are used below, but we have also applied our protocol to arrays of $N=9$ atoms~\cite{SI}. Each atom can occupy a ground state $|g\rangle$ or an excited (Rydberg) state $|r\rangle$, and two atoms excited to the Rydberg state at a distance $r$ interact with a van der Waals-type potential $V(r) \propto r^{-6}$.
When subjected to a uniform laser drive, the effective Hamiltonian of the many-body system can be written as~\cite{Bernien2017, Pohl2010, vanBinjen2011, Schauss2015}
\begin{equation}
\hat H(\Omega, \Delta) = -\Delta \sum_{i=1}^N \hat n_i - \frac{\Omega}{2} \sum_{i=1}^N \hat \sigma_i^x + \sum_{i<j} \frac{V_{nn}}{|i-j|^6} \hat n_i \hat n_j,\label{eqn:Hryd}
\end{equation}
where $V_{nn}$ is the interaction strength between Rydberg atoms at adjacent sites, $\hat \sigma_i^\alpha$, with $\alpha=x,y,z$,
are the Pauli pseudo-spin operators at site $i$ (defined as $\hat \sigma_i^z = \ket{r_i}\bra{r_i} - \ket{g_i}\bra{g_i}$,
$\hat \sigma_i^x = \ket{r_i}\bra{g_i} + h.c.$, etc), and $\hat n_i = \frac{1}{2} \left( 1 + \hat \sigma_i^z \right)$ projects
onto the Rydberg state at site $i$.
The parameters $\Omega, \Delta$ denote the effective Rabi frequency and detuning, respectively, which characterize the laser drive, and can be
varied in time as $\Omega(t), \Delta(t)$ to drive the system into nontrivial ordered phases~\cite{Pohl2010,Bernien2017, Weimer2010, Sela2011}. 

We focus on the transition into the $\mathbb{Z}_2$ phase~\cite{Bernien2017}, where a high density of Rydberg excitations is energetically favorable, subject to the constraint that no two adjacent atoms are excited. The atoms are initially pumped into the fiducial
state $|g\,g\,g\,g\,g\dots\rangle$, coinciding with the ground state of Hamiltonian (\ref{eqn:Hryd}) at $t=0$. They then evolve adiabatically under a ``sweep'' of the laser parameters $\Omega(t), \Delta(t)$ for a time $T_{\textnormal{ev}}$,
with $\hat H(\Omega(T_{\textnormal{ev}}), \Delta(T_{\textnormal{ev}}))$ lying deep in the $\mathbb{Z}_2$ phase (Fig.~\ref{Fig::experiment}c). 
For our eight-atom system, the final $\mathbb{Z}_2$-ordered state at $t = T_{\textnormal{ev}}$ is well approximated by the ground state of the Rydberg Hamiltonian with a small transverse field and short-range interactions only~\cite{SI}:
 \begin{equation}
  \ket{\psi} = \frac{1}{\sqrt 2} \ket{r\:g\:r\:g\:g\:r\:g\:r\:} + \frac{1}{2} \ket{r\:g\:r\:g\:r\:g\:g\:r\:} +
\frac{1}{2} \ket{r\:g\:g\:r\:g\:r\:g\:r\:} .
\label{eq:PsiGS}
\end{equation}  
This state exhibits quantum fluctuations on two pairs of adjacent atoms, as indicated in Fig.~\ref{Fig::experiment}b. 

\paragraph{ Pure state ansatz.} The ground state of the Hamiltonian (\ref{eqn:Hryd}) has real, positive amplitudes in the occupation number basis $|\bm{\sigma}\rangle=|\sigma_1, \dots, \sigma_N\rangle$ -- defined as the simultaneous eigenstates of $\hat n_1, \dots, \hat n_N$ -- as long as $\Omega>0$~\cite{Bravyi2008}, which can always be arranged by applying a suitable global unitary~\footnote{The exact phases required to render the Rydberg Hamiltonian in the form (\ref{eqn:Hryd}) will vary between different experimental realizations, but as the final measurements are always taken in the occupation number basis this has no effect on observables, provided no variation in the laser phase occurs during evolution.}. Therefore, if the quantum state of the simulator evolves perfectly adiabatically and with negligible loss of purity, it is uniquely characterized by its probability distribution $p(\bsig)$ over projective measurements in the $|\bm{\sigma}\rangle$ basis, and at any time may be written as the pure state 
\begin{equation}
\ket{\psi} =\sum_{\bsig} \sqrt{p(\bsig)} \ket{\bsig}.
\label{eqn:pure-state-ansatz}
\end{equation}
 Of course, some loss of purity is inevitable -- in our experiments, due primarily to single-atom decay and dephasing processes~\cite{Levine2018} -- and the true state is described by a mixed density operator $\hat \rho$. Although this pure state approximation cannot capture all of the physics of the experimental state, it can in principle accurately describe local subsystems, to the extent that the corresponding reduced density operators of the true and reconstructed states agree~\cite{SI}. We adopt the pure, positive state ansatz in all of our reconstruction efforts below.

\begin{figure}[t]
\noindent \centering \includegraphics[width=\columnwidth]{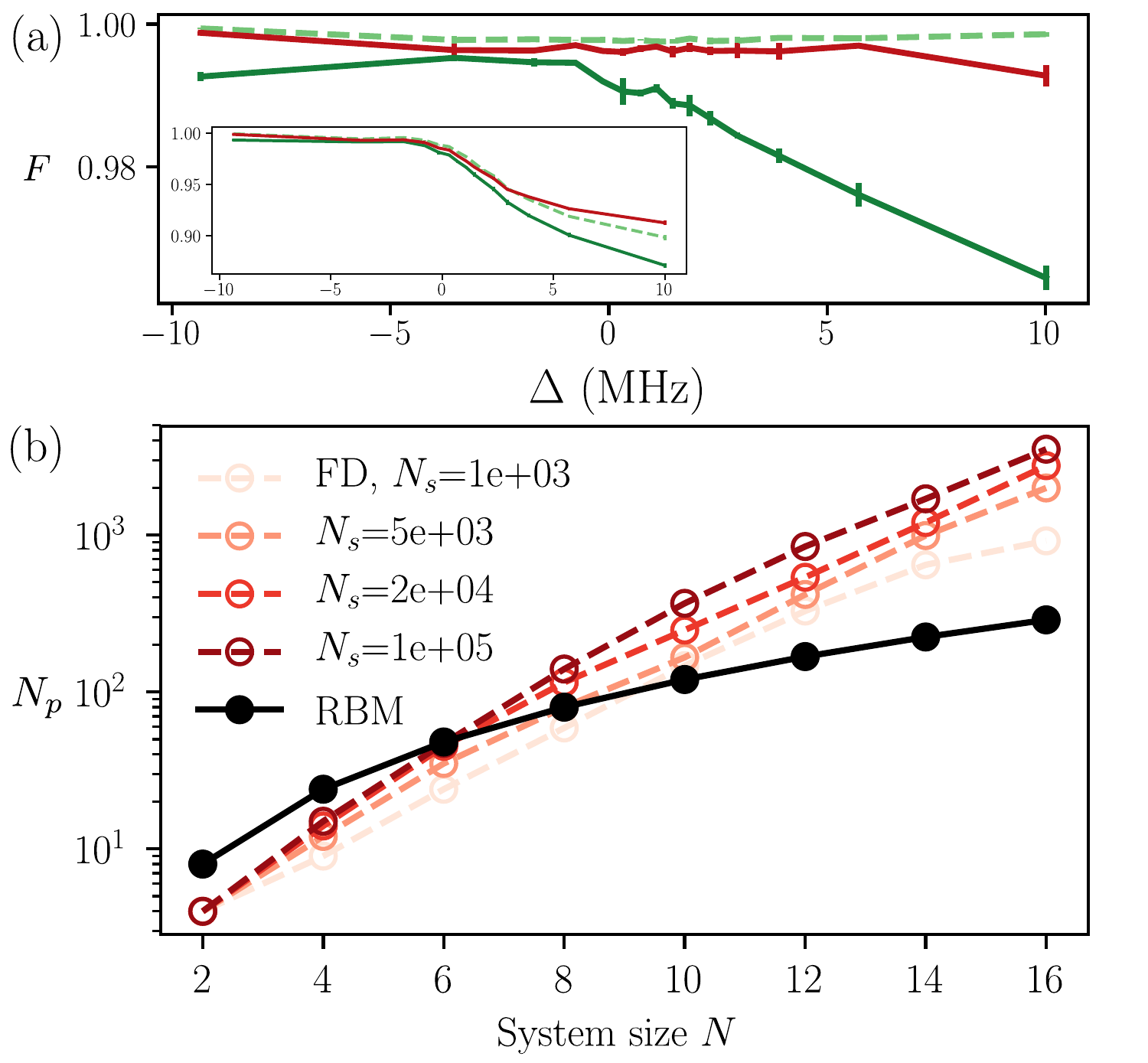}
\caption{Benchmarking RBM reconstruction. (a) Fidelity of reconstruction. We sample synthetic datasets from states obtained by exact time-evolution under the Hamiltonian~(\ref{eqn:Hryd}) without decoherence. The exact quantum state fidelity $F$ between the true state $\hat \rho$ and the reconstruction $\hat \rho_{\blda} = \ket{\psi_{\blda}} \bra{\psi_{\blda}}$ is plotted as a function of detuning $\Delta$. Training standard RBMs on datasets without measurement noise (green dashed line), we achieve uniformly high fidelities, demonstrating that the RBM wavefunction ansatz is capable of representing states relevant to our experiment. Training on datasets with measurement noise with (red solid line) and without (green solid line) noise-layer regularization shows how the modified training improves reconstruction. Inset: same data, for time-evolution including a realistic decoherence model.
		(b) Model size. Here we compare the number of parameters $N_p$ required to specify an RBM wavefunction with $N$ hidden units with the size of the frequency-distribution (FD) model required to perform direct inference (i.e. number of different configurations in the dataset), for a typical Rydberg ground state, as a function of system size $N$ and for several dataset sizes $N_s$. Note that the FD model size depends on $N_s$, while the RBM size does not.  For further discussion, see~\cite{SI}.}
\label{fig:benchmarking}
\end{figure}

\paragraph{Neural network model.}
While the quantum state~(\ref{eqn:pure-state-ansatz}) can in principle be inferred directly from a set of raw measurements (i.e. by inverting the measurement counts of each configuration to estimate $p(\bsig)$), such an approach is limited to small systems and very large datasets. In contrast, {\it generative models} used in unsupervised machine learning tasks can capture the structure of the distribution $p(\bsig)$, generalizing beyond a limited set of training samples. This results in a higher-fidelity reconstruction and a model size scaling polynomially in the system size (Fig.~\ref{fig:benchmarking}). Moreover, using a generative model rather than direct inference from the data enables automatic correction of this distribution for known measurement errors using a ``noise layer'' (see Fig.~\ref{Fig::experiment}a and description below). 

 We parametrize $p(\bsig)$ with a generative model known as an RBM~\cite{Ackley85,Smolensky1986}, a stochastic neural network with two layers of binary units. The ``visible'' layer $\bm{\sigma}$ describes the atomic states of the Rydberg chain in the occupation number basis, while a hidden layer $\bm{h}$ captures correlations between visible units. The RBM defines the following probability distribution for the visible layer:
\begin{align}
\pt (\bm{\sigma}) &= \frac{1}{Z_{\bm{\lambda}}} \sum_{\bh}  e^{\:\bm{h}^\top \bm{W} \bm{\sigma} + \bm{b} \cdot \bm{\sigma} + \bm{c} \cdot \bh}, \label{eqn:rbm} 
\end{align}
where $Z_{\bm{\lambda}}$ is a normalization constant, and the real-valued network parameters are
$\bm{\lambda}=\{\bm{W},\bm{b},\bm{c}\}$, with $\bm{W}$ being the weights connecting the two layers and $\bm{b}$ ($\bm{c}$) the visible (hidden) bias vectors. We use the visible layer of the RBM to define the projective measurement distribution $p(\bsig)$ of the pure state (\ref{eqn:pure-state-ansatz}), resulting in an RBM wavefunction with positive amplitudes~\cite{PhysRevB.94.165134}: $\psi_{\bm{\lambda}}(\bsig) = \langle \bsig | \psi_{\bm{\lambda}} \rangle = \sqrt{ \pt (\bsig)}$. We have numerically verified that this RBM wavefunction can accurately describe states relevant to our experiment, with a number of parameters scaling only quadratically in system size (Fig.~\ref{fig:benchmarking} and~\cite{SI}, Sec.~IV), in accordance with recent scaling studies for quantum Ising ground states~\cite{Sehayek2019}.
 We point out that, although pure states with nontrivial phases~\citep{torlai_2018_nnqst,carleo_2017_solving_mb}, as well as mixed state models~\citep{torlai_2018_ndo,Carrasquilla2018}, could be applied using similar neural network models, measurements in other bases %-- not used for the reconstruction process described in this Letter -- 
 would be required. % for this to provide an advantage.  

\begin{figure}[t]
\noindent \centering{} \includegraphics[width=\columnwidth]{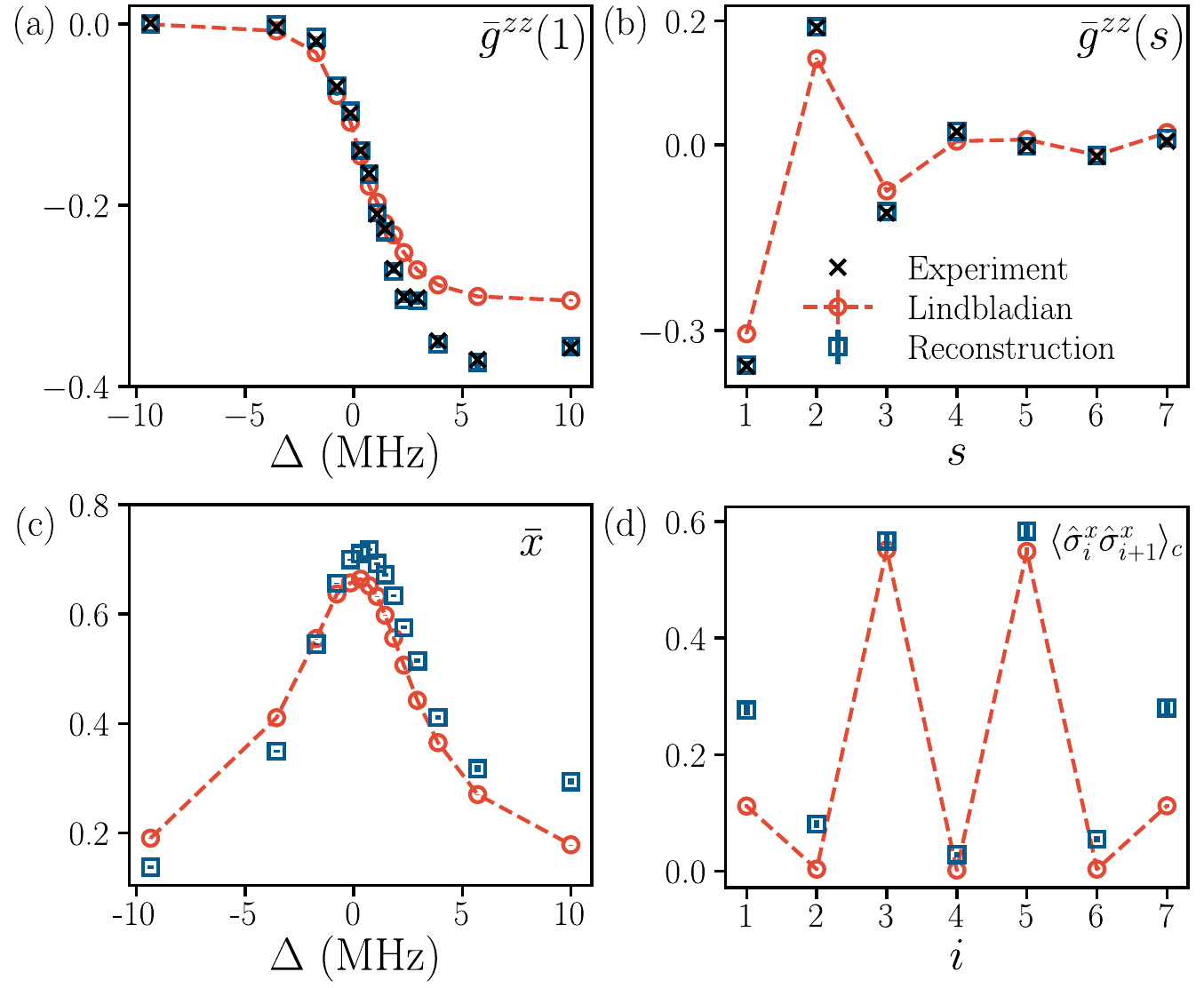}
\caption{{Few-body observables}. Comparison of the RBM reconstruction (squares) with the experiment results (crosses) and the predictions from the Lindbladian master equation (circles)~\cite{SI}. In order to facilitate comparison with experiment, the values reported in (a) and (b) for the RBM and Lindbladian observables are computed including the known measurement error rates $p(0|1) = 0.04$, $p(1|0) = 0.01$. (a) Nearest-neighbor correlations $\bar{g}^{zz}(1)$ in the $z$ basis, spatially averaged (see text for definition). (b) Average correlation $\bar{g}^{zz}(s)$ as a function of distance $s$ for $\Delta=10$ MHz. (c) Spatial average $\bar x$ of the transverse field $\langle \hat \sigma^x_i \rangle$. (d) Nearest-neighbor correlation $\langle \hat \sigma_i^x \hat \sigma_{i+1}^x \rangle_c$ as a function of position $i$ for $\Delta=10$ MHz. The two peaks correspond to the bonds highlighted in Fig.~\ref{Fig::experiment}b.}
\label{fig:obs}
\end{figure}

\paragraph{Measurement process and noise layer.} Measurement data consists of a collection of $N$-bit strings $\bm{\tau}=(\tau_1,\dots,\tau_N$), with $\tau_j=0,1$ indicating that atom $j$ was recorded as being in the ground $\ket{g}$ or Rydberg state $\ket{r}$ respectively~\cite{Bernien2017}. Such measurements are never perfect, and there are small measurement error probabilities $p(1|0)\sim1\%$, $p(0|1)\sim4\%$~\cite{Levine2018} for an atom in the ground state to be recorded as excited and vice-versa. These result in experimental data $\btau$ that do not correspond to projective measurements. Instead, the measurement process can be described as a positive-operator valued measure (POVM)~\cite{Nielsen2011} with measurement operators $\hat \Pi_{\btau}  = \sum_{\bsig} p(\btau | \bsig) \ket{\bsig}\bra{\bsig}$, where $p(\btau| \bsig) = \prod_{j=1}^N p(\tau_j| \sigma_j)$ is the probability of the experimentalist recording $\btau$ if the atoms are prepared in the state $\ket{\bsig}$. The probability distribution sampled in the experiment is then $P_{\textnormal{exp}}(\bm{\tau}) = \tr \left[ \hat \rho \hat \Pi_{\btau} \right]$. 

The experimental measurement process is incorporated into our model via a third binary layer, the so-called {\it noise layer} (Fig.~\ref{Fig::experiment}a), which represents the observed POVM outcomes $\btau$. The measurement error rates $p(\bm{\tau} | \bm{\sigma})$ are included as connections between the visible and noise layers~\cite{YichuanTang2012},  by assigning a probability $\pc (\bm{\btau}) = \sum_{\bm{\sigma}} p(\btau| \bsig) \pt (\bm{\sigma})$ to the measurement result $\btau$.

The full three-layer network is trained to learn parameters $\bm{\lambda}$ which maximize the log-likelihood of the recorded POVM outcomes under $\pc (\bm{\tau})$. During training, the noise layer prevents the parameters ${\blda}$ from fitting to spurious features in the data produced by measurement errors. This {\it noise layer regularization} significantly improves the fidelity between $\ket{\psi_{\bm{\lambda}}}$ and the state $\hat \rho$ underlying the data; numerical tests (Fig.~\ref{fig:benchmarking})  based on Lindbladian simulation of our experiment result in fidelities greater than 90\% for the full many-body state at the end of the sweep, even when decoherence processes are included. All reconstructions presented below are obtained in this fashion.

\begin{figure}[t]
\noindent \centering \includegraphics[width=\columnwidth]{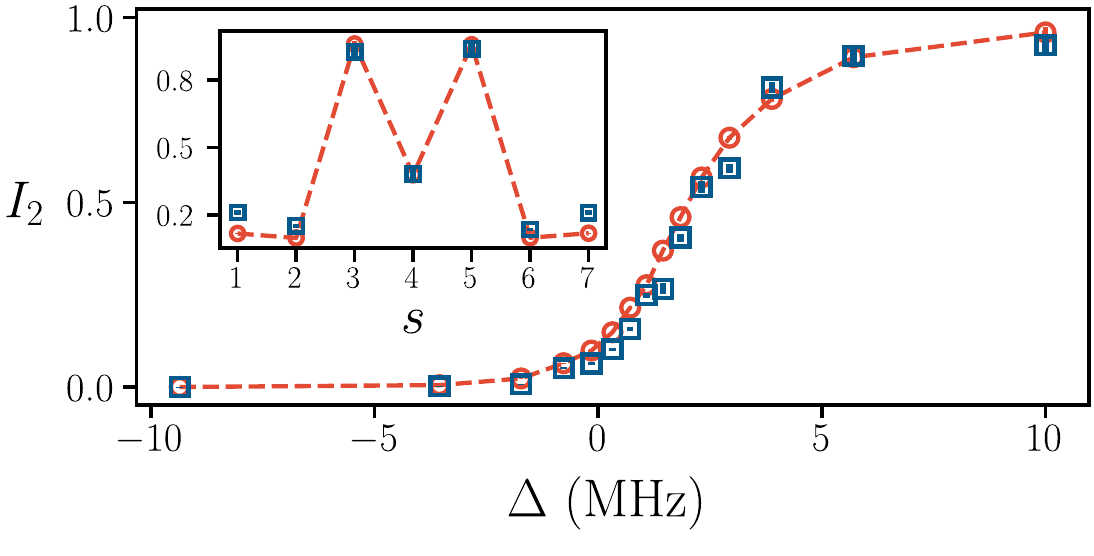}
\caption{{ R\'enyi Mutual Information}. The quantum (R\'enyi) mutual information $I_{2}$, defined as $I_2(s) = S_2(\hat \rho^A_s) + S_2(\hat \rho^B_s) - S_2(\hat \rho)$, where $S_2(\hat \rho) = -\log \tr \hat \rho^2$ is the second-order R\'enyi entropy, $\hat \rho$ is the (mixed) state of the whole system, and $\hat \rho^A_s, \hat \rho^B_s$ are the reduced density matrices for the subsystems $A_s = \{1, ..., s\}$, $B_s = \{ s+1, ... N \}$ respectively, defined by a partitioning of the system at bond $(s,s+1)$.
The mutual information is plotted for a partition at bond (3,4), as a function of detuning. Inset: The mutual information $I_{2}(s)$ as a function of the cut bond $s$ for $\Delta=10$ MHz.
}
\label{fig:entanglement}
\end{figure}

\paragraph{Experimental reconstruction.} In the experiment, at fifteen subsequent time-steps $t$ (Fig.~\ref{Fig::experiment}c), the sweep is halted and measurements $\btau$ are sampled from the state $\hat \rho(t)$. At each time-step, a dataset of around 3,000 samples is collected and used to train a three-layer model with $2N = 16$ hidden units. After training the networks, standard sampling methods can be applied to compute expectation values of observables, with a computational cost scaling polynomially in the network size~\cite{SI}. We consider in particular the connected correlation functions $\langle \hat \sigma_i^\alpha \hat \sigma_j^\alpha \rangle_c = \langle \hat \sigma_i^\alpha \hat \sigma_j^\alpha \rangle - \langle \hat \sigma_i^\alpha \rangle \langle \hat \sigma_j^\alpha \rangle$
for $\alpha = x, y, z$, and their spatial averages, $\bar{g}^{\alpha \alpha}(s) = \frac{1}{N-s} \sum_{i=1}^{N-s} \langle \hat \sigma_i^\alpha \hat \sigma_{i+s}^\alpha \rangle_c$.  

In Fig.~\ref{fig:obs}a-b, we verify that our reconstructions learn to represent their training sets, by examining their ability to accurately reproduce observables which are diagonal in the occupation number basis. The networks learn the strong two-body correlations $\langle \hat \sigma_i^z \hat \sigma_j^z \rangle_c$ present in the experimental data. We compare the results of the reconstruction process to the exact solutions of a Lindblad master equation for the full many-body evolution. Our Lindbladian simulation predicts Rydberg excitation profiles in excellent agreement with experiment, but its significantly weaker correlations suggest our model for the sweep dynamics is partially incomplete.

Turning to experimentally inaccessible quantities (Fig.~\ref{fig:obs}c-d), the reconstructions and simulation agree qualitatively in the temporal and spatial profiles of the transverse field $\langle \hat \sigma_i^x \rangle$ and its two-point correlation function, although the RBMs predict somewhat larger values in the ordered phase. Note that the distinct spatial variation of the transverse field correlations, a signature of quantum fluctuations captured in the approximate state~(\ref{eq:PsiGS}), is reconstructed directly from our experimental data. Training on synthetic data~\cite{SI} indicates that a large portion of the disagreement between reconstruction and simulation is due to the discrepancy between our Lindbladian model and experiment evident in Fig.~\ref{fig:obs}a-b, not the RBM model itself.

Beyond few-body observables, an important question is whether entanglement properties are reproduced accurately in reconstruction.  
From our RBMs, the R\'enyi entropy -- which requires specialized or hardware-specific protocols to access directly in experiment~\cite{Islam2015, Brydges2018} -- may be extracted in a scalable fashion by applying a state-replication and swap procedure virtually~\cite{Hastings2010,torlai_2018_nnqst}. In fact, for pure experimental states, positive-pure ansatzes such as the RBM wavefunction provide a lower bound on the mutual information defined by the R\'enyi entropy (\cite{Zhang2011}, \cite{Grover2015}, see also \cite{SI}, Sec. VIII), regardless of the sign structure of the true state. We demonstrate a reconstruction of the mutual information defined by the R\'enyi entropy in Fig.~\ref{fig:entanglement}, finding that the RBM values are in remarkable agreement with the results of numerical simulation. Reconstructions on experimental states of $N=9$ capture a similar buildup in the mutual information during the sweep predicted by Lindbladian simulation~\cite{SI}.

\paragraph{Conclusions.} 
In this Letter, we have demonstrated neural-network reconstruction of experimental quantum states from data produced by a programmable Rydberg-atom quantum simulator. By leveraging the real-positive nature of the ground state wavefunction expected from the effective Hamiltonian, we trained restricted Boltzmann machines on measurements in the occupation basis only. 
An additional noise layer was added to the standard RBM architecture to mitigate measurement errors.
Once trained, the RBM was queried to produce a variety of observables not accessible in the original experimental setup, including the R\'enyi entropy - a basis independent measure of the quantum entanglement of the wavefunction. 

Our approach can be integrated without alteration into existing platforms where a positive wavefunction ansatz is a valid approximation, such as Bose-Hubbard experiments and some non-frustrated quantum spin simulators~\cite{Bakr2009, Weitenberg2011,  Kaufman2016, Labuhn2016}. 
Access to multiple measurement bases would allow enhanced certification of the reconstruction, by providing direct experimental access to observables which are informationally complete for local subsystems.
Also, with access to multiple bases the RBM protocol can be easily adapted to reconstruct non-positive and complex wavefunctions~\cite{torlai_2018_nnqst}. Identifying the minimal set of measurement bases and the optimal protocol to collect the statistics represents a crucial step towards reconstruction of quantum states prepared by fermionic quantum simulators and non-equilibrium dynamics~\cite{Cheuk2015, Greif2016}. 

In conclusion, machine learning techniques offer a means of increasing the amount of useful information that can be extracted from experiments, especially when hardware constrains the quantity or quality of accessible measurements. They can be used to offload the burden of technically expensive -- or fundamentally impossible -- measurements from experimental platforms in a noise-resilient fashion. We expect experimentalists will profit from deeper integration of machine learning architectures with quantum devices.

\begin{acknowledgments}
We thank Dmitry Abanin for helpful discussions, and Soonwon Choi and Hannes Pichler for pointing out the bound on R\'enyi entropies. M.E. and B.T. acknowledge funding provided by the Institute for Quantum Information and Matter, an NSF Physics Frontiers Center (NSF Grant PHY-1733907), as well as the NSF CAREER award (1753386), and the AFOSR YIP (FA9550-19-1-0044). The Flatiron Institute is supported by the Simons Foundation. R.G.M. was supported by NSERC of Canada, a Canada Research Chair, and the Perimeter Institute for Theoretical Physics. Research at Perimeter Institute is supported through Industry Canada and by the Province of Ontario through the Ministry of Research \& Innovation. This research was supported in part by NSF Grant No. PHY-1748958, NIH Grant No. R25GM067110, and the Gordon and Betty Moore Foundation Grant No. 2919.01.
\end{acknowledgments}

\bibliographystyle{apsrev4-1}
\bibliography{bibliography.bib}

\begin{thebibliography}{66}%
\makeatletter
\providecommand \@ifxundefined [1]{%
 \@ifx{#1\undefined}
}%
\providecommand \@ifnum [1]{%
 \ifnum #1\expandafter \@firstoftwo
 \else \expandafter \@secondoftwo
 \fi
}%
\providecommand \@ifx [1]{%
 \ifx #1\expandafter \@firstoftwo
 \else \expandafter \@secondoftwo
 \fi
}%
\providecommand \natexlab [1]{#1}%
\providecommand \enquote  [1]{``#1''}%
\providecommand \bibnamefont  [1]{#1}%
\providecommand \bibfnamefont [1]{#1}%
\providecommand \citenamefont [1]{#1}%
\providecommand \href@noop [0]{\@secondoftwo}%
\providecommand \href [0]{\begingroup \@sanitize@url \@href}%
\providecommand \@href[1]{\@@startlink{#1}\@@href}%
\providecommand \@@href[1]{\endgroup#1\@@endlink}%
\providecommand \@sanitize@url [0]{\catcode `\\12\catcode `\$12\catcode
  `\&12\catcode `\#12\catcode `\^12\catcode `\_12\catcode `\%12\relax}%
\providecommand \@@startlink[1]{}%
\providecommand \@@endlink[0]{}%
\providecommand \url  [0]{\begingroup\@sanitize@url \@url }%
\providecommand \@url [1]{\endgroup\@href {#1}{\urlprefix }}%
\providecommand \urlprefix  [0]{URL }%
\providecommand \Eprint [0]{\href }%
\providecommand \doibase [0]{http://dx.doi.org/}%
\providecommand \selectlanguage [0]{\@gobble}%
\providecommand \bibinfo  [0]{\@secondoftwo}%
\providecommand \bibfield  [0]{\@secondoftwo}%
\providecommand \translation [1]{[#1]}%
\providecommand \BibitemOpen [0]{}%
\providecommand \bibitemStop [0]{}%
\providecommand \bibitemNoStop [0]{.\EOS\space}%
\providecommand \EOS [0]{\spacefactor3000\relax}%
\providecommand \BibitemShut  [1]{\csname bibitem#1\endcsname}%
\let\auto@bib@innerbib\@empty
%</preamble>
\bibitem [{\citenamefont {Banaszek}\ \emph {et~al.}(2013)\citenamefont
  {Banaszek}, \citenamefont {Cramer},\ and\ \citenamefont
  {Gross}}]{Banaszek2013}%
  \BibitemOpen
  \bibfield  {author} {\bibinfo {author} {\bibfnamefont {K.}~\bibnamefont
  {Banaszek}}, \bibinfo {author} {\bibfnamefont {M.}~\bibnamefont {Cramer}}, \
  and\ \bibinfo {author} {\bibfnamefont {D.}~\bibnamefont {Gross}},\ }\href
  {\doibase 10.1088/1367-2630/15/12/125020} {\bibfield  {journal} {\bibinfo
  {journal} {New Journal of Physics}\ }\textbf {\bibinfo {volume} {15}},\
  \bibinfo {pages} {125020} (\bibinfo {year} {2013})}\BibitemShut {NoStop}%
\bibitem [{\citenamefont {H{\"{a}}ffner}\ \emph {et~al.}(2005)\citenamefont
  {H{\"{a}}ffner}, \citenamefont {H{\"{a}}nsel}, \citenamefont {Roos},
  \citenamefont {Benhelm}, \citenamefont {Chek-al kar}, \citenamefont
  {Chwalla}, \citenamefont {K{\"{o}}rber}, \citenamefont {Rapol}, \citenamefont
  {Riebe}, \citenamefont {Schmidt}, \citenamefont {Becher}, \citenamefont
  {G{\"{u}}hne}, \citenamefont {D{\"{u}}r},\ and\ \citenamefont
  {Blatt}}]{Haffner2005}%
  \BibitemOpen
  \bibfield  {author} {\bibinfo {author} {\bibfnamefont {H.}~\bibnamefont
  {H{\"{a}}ffner}}, \bibinfo {author} {\bibfnamefont {W.}~\bibnamefont
  {H{\"{a}}nsel}}, \bibinfo {author} {\bibfnamefont {C.~F.}\ \bibnamefont
  {Roos}}, \bibinfo {author} {\bibfnamefont {J.}~\bibnamefont {Benhelm}},
  \bibinfo {author} {\bibfnamefont {D.}~\bibnamefont {Chek-al kar}}, \bibinfo
  {author} {\bibfnamefont {M.}~\bibnamefont {Chwalla}}, \bibinfo {author}
  {\bibfnamefont {T.}~\bibnamefont {K{\"{o}}rber}}, \bibinfo {author}
  {\bibfnamefont {U.~D.}\ \bibnamefont {Rapol}}, \bibinfo {author}
  {\bibfnamefont {M.}~\bibnamefont {Riebe}}, \bibinfo {author} {\bibfnamefont
  {P.~O.}\ \bibnamefont {Schmidt}}, \bibinfo {author} {\bibfnamefont
  {C.}~\bibnamefont {Becher}}, \bibinfo {author} {\bibfnamefont
  {O.}~\bibnamefont {G{\"{u}}hne}}, \bibinfo {author} {\bibfnamefont
  {W.}~\bibnamefont {D{\"{u}}r}}, \ and\ \bibinfo {author} {\bibfnamefont
  {R.}~\bibnamefont {Blatt}},\ }\href {\doibase 10.1038/nature04279} {\bibfield
   {journal} {\bibinfo  {journal} {Nature}\ }\textbf {\bibinfo {volume}
  {438}},\ \bibinfo {pages} {643} (\bibinfo {year} {2005})}\BibitemShut
  {NoStop}%
\bibitem [{\citenamefont {Cramer}\ \emph {et~al.}(2010)\citenamefont {Cramer},
  \citenamefont {Plenio}, \citenamefont {Flammia}, \citenamefont {Somma},
  \citenamefont {Gross}, \citenamefont {Bartlett}, \citenamefont
  {Landon-Cardinal}, \citenamefont {Poulin},\ and\ \citenamefont
  {Liu}}]{Cramer2010}%
  \BibitemOpen
  \bibfield  {author} {\bibinfo {author} {\bibfnamefont {M.}~\bibnamefont
  {Cramer}}, \bibinfo {author} {\bibfnamefont {M.~B.}\ \bibnamefont {Plenio}},
  \bibinfo {author} {\bibfnamefont {S.~T.}\ \bibnamefont {Flammia}}, \bibinfo
  {author} {\bibfnamefont {R.}~\bibnamefont {Somma}}, \bibinfo {author}
  {\bibfnamefont {D.}~\bibnamefont {Gross}}, \bibinfo {author} {\bibfnamefont
  {S.~D.}\ \bibnamefont {Bartlett}}, \bibinfo {author} {\bibfnamefont
  {O.}~\bibnamefont {Landon-Cardinal}}, \bibinfo {author} {\bibfnamefont
  {D.}~\bibnamefont {Poulin}}, \ and\ \bibinfo {author} {\bibfnamefont {Y.-K.}\
  \bibnamefont {Liu}},\ }\href {\doibase 10.1038/ncomms1147} {\bibfield
  {journal} {\bibinfo  {journal} {Nature Communications}\ }\textbf {\bibinfo
  {volume} {1}},\ \bibinfo {pages} {149} (\bibinfo {year} {2010})}\BibitemShut
  {NoStop}%
\bibitem [{\citenamefont {Lee}\ and\ \citenamefont
  {Landon-Cardinal}(2015)}]{Lee2015}%
  \BibitemOpen
  \bibfield  {author} {\bibinfo {author} {\bibfnamefont {J.~Y.}\ \bibnamefont
  {Lee}}\ and\ \bibinfo {author} {\bibfnamefont {O.}~\bibnamefont
  {Landon-Cardinal}},\ }\href {\doibase 10.1103/PhysRevA.91.062128} {\bibfield
  {journal} {\bibinfo  {journal} {Phys. Rev. A}\ }\textbf {\bibinfo {volume}
  {91}},\ \bibinfo {pages} {062128} (\bibinfo {year} {2015})}\BibitemShut
  {NoStop}%
\bibitem [{\citenamefont {Riofr{\'{i}}o}\ \emph {et~al.}(2017)\citenamefont
  {Riofr{\'{i}}o}, \citenamefont {Gross}, \citenamefont {Flammia},
  \citenamefont {Monz}, \citenamefont {Nigg}, \citenamefont {Blatt},\ and\
  \citenamefont {Eisert}}]{Riofrio2017}%
  \BibitemOpen
  \bibfield  {author} {\bibinfo {author} {\bibfnamefont {C.~A.}\ \bibnamefont
  {Riofr{\'{i}}o}}, \bibinfo {author} {\bibfnamefont {D.}~\bibnamefont
  {Gross}}, \bibinfo {author} {\bibfnamefont {S.~T.}\ \bibnamefont {Flammia}},
  \bibinfo {author} {\bibfnamefont {T.}~\bibnamefont {Monz}}, \bibinfo {author}
  {\bibfnamefont {D.}~\bibnamefont {Nigg}}, \bibinfo {author} {\bibfnamefont
  {R.}~\bibnamefont {Blatt}}, \ and\ \bibinfo {author} {\bibfnamefont
  {J.}~\bibnamefont {Eisert}},\ }\href {\doibase 10.1038/ncomms15305}
  {\bibfield  {journal} {\bibinfo  {journal} {Nature Communications}\ }\textbf
  {\bibinfo {volume} {8}},\ \bibinfo {pages} {15305} (\bibinfo {year}
  {2017})}\BibitemShut {NoStop}%
\bibitem [{\citenamefont {Lanyon}\ \emph {et~al.}(2017)\citenamefont {Lanyon},
  \citenamefont {Maier}, \citenamefont {Holz{\"{a}}pfel}, \citenamefont
  {Baumgratz}, \citenamefont {Hempel}, \citenamefont {Jurcevic}, \citenamefont
  {Dhand}, \citenamefont {Buyskikh}, \citenamefont {Daley}, \citenamefont
  {Cramer}, \citenamefont {Plenio}, \citenamefont {Blatt},\ and\ \citenamefont
  {Roos}}]{Lanyon2017}%
  \BibitemOpen
  \bibfield  {author} {\bibinfo {author} {\bibfnamefont {B.~P.}\ \bibnamefont
  {Lanyon}}, \bibinfo {author} {\bibfnamefont {C.}~\bibnamefont {Maier}},
  \bibinfo {author} {\bibfnamefont {M.}~\bibnamefont {Holz{\"{a}}pfel}},
  \bibinfo {author} {\bibfnamefont {T.}~\bibnamefont {Baumgratz}}, \bibinfo
  {author} {\bibfnamefont {C.}~\bibnamefont {Hempel}}, \bibinfo {author}
  {\bibfnamefont {P.}~\bibnamefont {Jurcevic}}, \bibinfo {author}
  {\bibfnamefont {I.}~\bibnamefont {Dhand}}, \bibinfo {author} {\bibfnamefont
  {A.~S.}\ \bibnamefont {Buyskikh}}, \bibinfo {author} {\bibfnamefont {A.~J.}\
  \bibnamefont {Daley}}, \bibinfo {author} {\bibfnamefont {M.}~\bibnamefont
  {Cramer}}, \bibinfo {author} {\bibfnamefont {M.~B.}\ \bibnamefont {Plenio}},
  \bibinfo {author} {\bibfnamefont {R.}~\bibnamefont {Blatt}}, \ and\ \bibinfo
  {author} {\bibfnamefont {C.~F.}\ \bibnamefont {Roos}},\ }\href {\doibase
  10.1038/nphys4244} {\bibfield  {journal} {\bibinfo  {journal} {Nature
  Physics}\ }\textbf {\bibinfo {volume} {13}},\ \bibinfo {pages} {1158}
  (\bibinfo {year} {2017})}\BibitemShut {NoStop}%
\bibitem [{\citenamefont {T\'oth}\ \emph {et~al.}(2010)\citenamefont {T\'oth},
  \citenamefont {Wieczorek}, \citenamefont {Gross}, \citenamefont {Krischek},
  \citenamefont {Schwemmer},\ and\ \citenamefont {Weinfurter}}]{Toth2010}%
  \BibitemOpen
  \bibfield  {author} {\bibinfo {author} {\bibfnamefont {G.}~\bibnamefont
  {T\'oth}}, \bibinfo {author} {\bibfnamefont {W.}~\bibnamefont {Wieczorek}},
  \bibinfo {author} {\bibfnamefont {D.}~\bibnamefont {Gross}}, \bibinfo
  {author} {\bibfnamefont {R.}~\bibnamefont {Krischek}}, \bibinfo {author}
  {\bibfnamefont {C.}~\bibnamefont {Schwemmer}}, \ and\ \bibinfo {author}
  {\bibfnamefont {H.}~\bibnamefont {Weinfurter}},\ }\href {\doibase
  10.1103/PhysRevLett.105.250403} {\bibfield  {journal} {\bibinfo  {journal}
  {Phys. Rev. Lett.}\ }\textbf {\bibinfo {volume} {105}},\ \bibinfo {pages}
  {250403} (\bibinfo {year} {2010})}\BibitemShut {NoStop}%
\bibitem [{\citenamefont {Je{\v{z}}ek}\ \emph {et~al.}(2003)\citenamefont
  {Je{\v{z}}ek}, \citenamefont {Fiur{\'{a}}{\v{s}}ek},\ and\ \citenamefont
  {Hradil}}]{Jezek2003}%
  \BibitemOpen
  \bibfield  {author} {\bibinfo {author} {\bibfnamefont {M.}~\bibnamefont
  {Je{\v{z}}ek}}, \bibinfo {author} {\bibfnamefont {J.}~\bibnamefont
  {Fiur{\'{a}}{\v{s}}ek}}, \ and\ \bibinfo {author} {\bibfnamefont
  {Z.}~\bibnamefont {Hradil}},\ }\href {\doibase 10.1103/PhysRevA.68.012305}
  {\bibfield  {journal} {\bibinfo  {journal} {Physical Review A}\ }\textbf
  {\bibinfo {volume} {68}},\ \bibinfo {pages} {012305} (\bibinfo {year}
  {2003})}\BibitemShut {NoStop}%
\bibitem [{\citenamefont {Preskill}(2018)}]{Preskill2018}%
  \BibitemOpen
  \bibfield  {author} {\bibinfo {author} {\bibfnamefont {J.}~\bibnamefont
  {Preskill}},\ }\href {\doibase 10.22331/q-2018-08-06-79} {\bibfield
  {journal} {\bibinfo  {journal} {{Quantum}}\ }\textbf {\bibinfo {volume}
  {2}},\ \bibinfo {pages} {79} (\bibinfo {year} {2018})}\BibitemShut {NoStop}%
\bibitem [{\citenamefont {Hinton}(2006)}]{Hinton2006AE}%
  \BibitemOpen
  \bibfield  {author} {\bibinfo {author} {\bibfnamefont {G.~E.}\ \bibnamefont
  {Hinton}},\ }\href {\doibase 10.1126/science.1127647} {\bibfield  {journal}
  {\bibinfo  {journal} {Science}\ }\textbf {\bibinfo {volume} {313}},\ \bibinfo
  {pages} {504} (\bibinfo {year} {2006})}\BibitemShut {NoStop}%
\bibitem [{\citenamefont {Graves}\ \emph {et~al.}(2013)\citenamefont {Graves},
  \citenamefont {Mohamed},\ and\ \citenamefont {Hinton}}]{Graves2013}%
  \BibitemOpen
  \bibfield  {author} {\bibinfo {author} {\bibfnamefont {A.}~\bibnamefont
  {Graves}}, \bibinfo {author} {\bibfnamefont {A.-r.}\ \bibnamefont {Mohamed}},
  \ and\ \bibinfo {author} {\bibfnamefont {G.}~\bibnamefont {Hinton}},\ }in\
  \href {\doibase 10.1109/ICASSP.2013.6638947} {\emph {\bibinfo {booktitle}
  {2013 IEEE International Conference on Acoustics, Speech and Signal
  Processing}}}\ (\bibinfo  {publisher} {IEEE},\ \bibinfo {year} {2013})\ pp.\
  \bibinfo {pages} {6645--6649}\BibitemShut {NoStop}%
\bibitem [{\citenamefont {LeCun}\ \emph {et~al.}(2015)\citenamefont {LeCun},
  \citenamefont {Bengio},\ and\ \citenamefont {Hinton}}]{LeCun2015}%
  \BibitemOpen
  \bibfield  {author} {\bibinfo {author} {\bibfnamefont {Y.}~\bibnamefont
  {LeCun}}, \bibinfo {author} {\bibfnamefont {Y.}~\bibnamefont {Bengio}}, \
  and\ \bibinfo {author} {\bibfnamefont {G.}~\bibnamefont {Hinton}},\ }\href
  {\doibase 10.1038/nature14539} {\bibfield  {journal} {\bibinfo  {journal}
  {Nature}\ }\textbf {\bibinfo {volume} {521}},\ \bibinfo {pages} {436}
  (\bibinfo {year} {2015})}\BibitemShut {NoStop}%
\bibitem [{\citenamefont {Rem}\ \emph {et~al.}(2019)\citenamefont {Rem},
  \citenamefont {K{\"a}ming}, \citenamefont {Tarnowski}, \citenamefont
  {Asteria}, \citenamefont {Fl{\"a}schner}, \citenamefont {Becker},
  \citenamefont {Sengstock},\ and\ \citenamefont {Weitenberg}}]{Rem2018}%
  \BibitemOpen
  \bibfield  {author} {\bibinfo {author} {\bibfnamefont {B.~S.}\ \bibnamefont
  {Rem}}, \bibinfo {author} {\bibfnamefont {N.}~\bibnamefont {K{\"a}ming}},
  \bibinfo {author} {\bibfnamefont {M.}~\bibnamefont {Tarnowski}}, \bibinfo
  {author} {\bibfnamefont {L.}~\bibnamefont {Asteria}}, \bibinfo {author}
  {\bibfnamefont {N.}~\bibnamefont {Fl{\"a}schner}}, \bibinfo {author}
  {\bibfnamefont {C.}~\bibnamefont {Becker}}, \bibinfo {author} {\bibfnamefont
  {K.}~\bibnamefont {Sengstock}}, \ and\ \bibinfo {author} {\bibfnamefont
  {C.}~\bibnamefont {Weitenberg}},\ }\href {\doibase 10.1038/s41567-019-0554-0}
  {\bibfield  {journal} {\bibinfo  {journal} {Nature Physics}\ }\textbf
  {\bibinfo {volume} {15}},\ \bibinfo {pages} {917} (\bibinfo {year}
  {2019})}\BibitemShut {NoStop}%
\bibitem [{\citenamefont {Bohrdt}\ \emph {et~al.}(2019)\citenamefont {Bohrdt},
  \citenamefont {Chiu}, \citenamefont {Ji}, \citenamefont {Xu}, \citenamefont
  {Greif}, \citenamefont {Greiner}, \citenamefont {Demler}, \citenamefont
  {Grusdt},\ and\ \citenamefont {Knap}}]{Bohrdt2018}%
  \BibitemOpen
  \bibfield  {author} {\bibinfo {author} {\bibfnamefont {A.}~\bibnamefont
  {Bohrdt}}, \bibinfo {author} {\bibfnamefont {C.~S.}\ \bibnamefont {Chiu}},
  \bibinfo {author} {\bibfnamefont {G.}~\bibnamefont {Ji}}, \bibinfo {author}
  {\bibfnamefont {M.}~\bibnamefont {Xu}}, \bibinfo {author} {\bibfnamefont
  {D.}~\bibnamefont {Greif}}, \bibinfo {author} {\bibfnamefont
  {M.}~\bibnamefont {Greiner}}, \bibinfo {author} {\bibfnamefont
  {E.}~\bibnamefont {Demler}}, \bibinfo {author} {\bibfnamefont
  {F.}~\bibnamefont {Grusdt}}, \ and\ \bibinfo {author} {\bibfnamefont
  {M.}~\bibnamefont {Knap}},\ }\href {\doibase 10.1038/s41567-019-0565-x}
  {\bibfield  {journal} {\bibinfo  {journal} {Nature Physics}\ }\textbf
  {\bibinfo {volume} {15}},\ \bibinfo {pages} {921} (\bibinfo {year}
  {2019})}\BibitemShut {NoStop}%
\bibitem [{\citenamefont {Seif}\ \emph {et~al.}(2018)\citenamefont {Seif},
  \citenamefont {Landsman}, \citenamefont {Linke}, \citenamefont {Figgatt},
  \citenamefont {Monroe},\ and\ \citenamefont {Hafezi}}]{Seif2018}%
  \BibitemOpen
  \bibfield  {author} {\bibinfo {author} {\bibfnamefont {A.}~\bibnamefont
  {Seif}}, \bibinfo {author} {\bibfnamefont {K.~A.}\ \bibnamefont {Landsman}},
  \bibinfo {author} {\bibfnamefont {N.~M.}\ \bibnamefont {Linke}}, \bibinfo
  {author} {\bibfnamefont {C.}~\bibnamefont {Figgatt}}, \bibinfo {author}
  {\bibfnamefont {C.}~\bibnamefont {Monroe}}, \ and\ \bibinfo {author}
  {\bibfnamefont {M.}~\bibnamefont {Hafezi}},\ }\href {\doibase
  10.1088/1361-6455/aad62b} {\bibfield  {journal} {\bibinfo  {journal} {Journal
  of Physics B: Atomic, Molecular and Optical Physics}\ }\textbf {\bibinfo
  {volume} {51}},\ \bibinfo {pages} {174006} (\bibinfo {year} {2018})},\
  \Eprint {http://arxiv.org/abs/1804.07718} {arXiv:1804.07718} \BibitemShut
  {NoStop}%
\bibitem [{\citenamefont {{Torlai}}\ and\ \citenamefont
  {{Melko}}(2019)}]{rbm_nisq}%
  \BibitemOpen
  \bibfield  {author} {\bibinfo {author} {\bibfnamefont {G.}~\bibnamefont
  {{Torlai}}}\ and\ \bibinfo {author} {\bibfnamefont {R.~G.}\ \bibnamefont
  {{Melko}}},\ }\href@noop {} {\bibfield  {journal} {\bibinfo  {journal} {arXiv
  e-prints}\ ,\ \bibinfo {eid} {arXiv:1905.04312}} (\bibinfo {year} {2019})},\
  \Eprint {http://arxiv.org/abs/1905.04312} {arXiv:1905.04312 [quant-ph]}
  \BibitemShut {NoStop}%
\bibitem [{\citenamefont {Bernien}\ \emph {et~al.}(2017)\citenamefont
  {Bernien}, \citenamefont {Schwartz}, \citenamefont {Keesling}, \citenamefont
  {Levine}, \citenamefont {Omran}, \citenamefont {Pichler}, \citenamefont
  {Choi}, \citenamefont {Zibrov}, \citenamefont {Endres}, \citenamefont
  {Greiner}, \citenamefont {Vuleti{\'{c}}},\ and\ \citenamefont
  {Lukin}}]{Bernien2017}%
  \BibitemOpen
  \bibfield  {author} {\bibinfo {author} {\bibfnamefont {H.}~\bibnamefont
  {Bernien}}, \bibinfo {author} {\bibfnamefont {S.}~\bibnamefont {Schwartz}},
  \bibinfo {author} {\bibfnamefont {A.}~\bibnamefont {Keesling}}, \bibinfo
  {author} {\bibfnamefont {H.}~\bibnamefont {Levine}}, \bibinfo {author}
  {\bibfnamefont {A.}~\bibnamefont {Omran}}, \bibinfo {author} {\bibfnamefont
  {H.}~\bibnamefont {Pichler}}, \bibinfo {author} {\bibfnamefont
  {S.}~\bibnamefont {Choi}}, \bibinfo {author} {\bibfnamefont {A.~S.}\
  \bibnamefont {Zibrov}}, \bibinfo {author} {\bibfnamefont {M.}~\bibnamefont
  {Endres}}, \bibinfo {author} {\bibfnamefont {M.}~\bibnamefont {Greiner}},
  \bibinfo {author} {\bibfnamefont {V.}~\bibnamefont {Vuleti{\'{c}}}}, \ and\
  \bibinfo {author} {\bibfnamefont {M.~D.}\ \bibnamefont {Lukin}},\ }\href
  {\doibase 10.1038/nature24622} {\bibfield  {journal} {\bibinfo  {journal}
  {Nature}\ }\textbf {\bibinfo {volume} {551}},\ \bibinfo {pages} {579}
  (\bibinfo {year} {2017})}\BibitemShut {NoStop}%
\bibitem [{\citenamefont {Endres}\ \emph {et~al.}(2016)\citenamefont {Endres},
  \citenamefont {Bernien}, \citenamefont {Keesling}, \citenamefont {Levine},
  \citenamefont {Anschuetz}, \citenamefont {Krajenbrink}, \citenamefont
  {Senko}, \citenamefont {Vuletic}, \citenamefont {Greiner},\ and\
  \citenamefont {Lukin}}]{Endres2016}%
  \BibitemOpen
  \bibfield  {author} {\bibinfo {author} {\bibfnamefont {M.}~\bibnamefont
  {Endres}}, \bibinfo {author} {\bibfnamefont {H.}~\bibnamefont {Bernien}},
  \bibinfo {author} {\bibfnamefont {A.}~\bibnamefont {Keesling}}, \bibinfo
  {author} {\bibfnamefont {H.}~\bibnamefont {Levine}}, \bibinfo {author}
  {\bibfnamefont {E.~R.}\ \bibnamefont {Anschuetz}}, \bibinfo {author}
  {\bibfnamefont {A.}~\bibnamefont {Krajenbrink}}, \bibinfo {author}
  {\bibfnamefont {C.}~\bibnamefont {Senko}}, \bibinfo {author} {\bibfnamefont
  {V.}~\bibnamefont {Vuletic}}, \bibinfo {author} {\bibfnamefont
  {M.}~\bibnamefont {Greiner}}, \ and\ \bibinfo {author} {\bibfnamefont
  {M.~D.}\ \bibnamefont {Lukin}},\ }\href {\doibase 10.1126/science.aah3752}
  {\bibfield  {journal} {\bibinfo  {journal} {Science}\ }\textbf {\bibinfo
  {volume} {354}},\ \bibinfo {pages} {1024} (\bibinfo {year}
  {2016})}\BibitemShut {NoStop}%
\bibitem [{\citenamefont {Schau{\ss}}\ \emph {et~al.}(2015)\citenamefont
  {Schau{\ss}}, \citenamefont {Zeiher}, \citenamefont {Fukuhara}, \citenamefont
  {Hild}, \citenamefont {Cheneau}, \citenamefont {Macr{\`{i}}}, \citenamefont
  {Pohl}, \citenamefont {Bloch},\ and\ \citenamefont {Gross}}]{Schauss2015}%
  \BibitemOpen
  \bibfield  {author} {\bibinfo {author} {\bibfnamefont {P.}~\bibnamefont
  {Schau{\ss}}}, \bibinfo {author} {\bibfnamefont {J.}~\bibnamefont {Zeiher}},
  \bibinfo {author} {\bibfnamefont {T.}~\bibnamefont {Fukuhara}}, \bibinfo
  {author} {\bibfnamefont {S.}~\bibnamefont {Hild}}, \bibinfo {author}
  {\bibfnamefont {M.}~\bibnamefont {Cheneau}}, \bibinfo {author} {\bibfnamefont
  {T.}~\bibnamefont {Macr{\`{i}}}}, \bibinfo {author} {\bibfnamefont
  {T.}~\bibnamefont {Pohl}}, \bibinfo {author} {\bibfnamefont {I.}~\bibnamefont
  {Bloch}}, \ and\ \bibinfo {author} {\bibfnamefont {C.}~\bibnamefont
  {Gross}},\ }\href {\doibase 10.1126/science.1258351} {\bibfield  {journal}
  {\bibinfo  {journal} {Science}\ }\textbf {\bibinfo {volume} {347}},\ \bibinfo
  {pages} {1455} (\bibinfo {year} {2015})}\BibitemShut {NoStop}%
\bibitem [{\citenamefont {Labuhn}\ \emph {et~al.}(2016)\citenamefont {Labuhn},
  \citenamefont {Barredo}, \citenamefont {Ravets}, \citenamefont
  {de~L{\'{e}}s{\'{e}}leuc}, \citenamefont {Macr{\`{i}}}, \citenamefont
  {Lahaye},\ and\ \citenamefont {Browaeys}}]{Labuhn2016}%
  \BibitemOpen
  \bibfield  {author} {\bibinfo {author} {\bibfnamefont {H.}~\bibnamefont
  {Labuhn}}, \bibinfo {author} {\bibfnamefont {D.}~\bibnamefont {Barredo}},
  \bibinfo {author} {\bibfnamefont {S.}~\bibnamefont {Ravets}}, \bibinfo
  {author} {\bibfnamefont {S.}~\bibnamefont {de~L{\'{e}}s{\'{e}}leuc}},
  \bibinfo {author} {\bibfnamefont {T.}~\bibnamefont {Macr{\`{i}}}}, \bibinfo
  {author} {\bibfnamefont {T.}~\bibnamefont {Lahaye}}, \ and\ \bibinfo {author}
  {\bibfnamefont {A.}~\bibnamefont {Browaeys}},\ }\href {\doibase
  10.1038/nature18274} {\bibfield  {journal} {\bibinfo  {journal} {Nature}\
  }\textbf {\bibinfo {volume} {534}},\ \bibinfo {pages} {667} (\bibinfo {year}
  {2016})}\BibitemShut {NoStop}%
\bibitem [{\citenamefont {Zeiher}\ \emph {et~al.}(2016)\citenamefont {Zeiher},
  \citenamefont {van Bijnen}, \citenamefont {Schau{\ss}}, \citenamefont {Hild},
  \citenamefont {Choi}, \citenamefont {Pohl}, \citenamefont {Bloch},\ and\
  \citenamefont {Gross}}]{Zeiher2016}%
  \BibitemOpen
  \bibfield  {author} {\bibinfo {author} {\bibfnamefont {J.}~\bibnamefont
  {Zeiher}}, \bibinfo {author} {\bibfnamefont {R.}~\bibnamefont {van Bijnen}},
  \bibinfo {author} {\bibfnamefont {P.}~\bibnamefont {Schau{\ss}}}, \bibinfo
  {author} {\bibfnamefont {S.}~\bibnamefont {Hild}}, \bibinfo {author}
  {\bibfnamefont {J.-y.}\ \bibnamefont {Choi}}, \bibinfo {author}
  {\bibfnamefont {T.}~\bibnamefont {Pohl}}, \bibinfo {author} {\bibfnamefont
  {I.}~\bibnamefont {Bloch}}, \ and\ \bibinfo {author} {\bibfnamefont
  {C.}~\bibnamefont {Gross}},\ }\href {\doibase 10.1038/nphys3835} {\bibfield
  {journal} {\bibinfo  {journal} {Nature Physics}\ }\textbf {\bibinfo {volume}
  {12}},\ \bibinfo {pages} {1095} (\bibinfo {year} {2016})}\BibitemShut
  {NoStop}%
\bibitem [{\citenamefont {Zeiher}\ \emph {et~al.}(2017)\citenamefont {Zeiher},
  \citenamefont {Choi}, \citenamefont {Rubio-Abadal}, \citenamefont {Pohl},
  \citenamefont {van Bijnen}, \citenamefont {Bloch},\ and\ \citenamefont
  {Gross}}]{Zeiher2017}%
  \BibitemOpen
  \bibfield  {author} {\bibinfo {author} {\bibfnamefont {J.}~\bibnamefont
  {Zeiher}}, \bibinfo {author} {\bibfnamefont {J.-y.}\ \bibnamefont {Choi}},
  \bibinfo {author} {\bibfnamefont {A.}~\bibnamefont {Rubio-Abadal}}, \bibinfo
  {author} {\bibfnamefont {T.}~\bibnamefont {Pohl}}, \bibinfo {author}
  {\bibfnamefont {R.}~\bibnamefont {van Bijnen}}, \bibinfo {author}
  {\bibfnamefont {I.}~\bibnamefont {Bloch}}, \ and\ \bibinfo {author}
  {\bibfnamefont {C.}~\bibnamefont {Gross}},\ }\href {\doibase
  10.1103/PhysRevX.7.041063} {\bibfield  {journal} {\bibinfo  {journal}
  {Physical Review X}\ }\textbf {\bibinfo {volume} {7}},\ \bibinfo {pages}
  {041063} (\bibinfo {year} {2017})}\BibitemShut {NoStop}%
\bibitem [{\citenamefont {Guardado-Sanchez}\ \emph {et~al.}(2018)\citenamefont
  {Guardado-Sanchez}, \citenamefont {Brown}, \citenamefont {Mitra},
  \citenamefont {Devakul}, \citenamefont {Huse}, \citenamefont {Schau{\ss}},\
  and\ \citenamefont {Bakr}}]{Guardado-Sanchez2018}%
  \BibitemOpen
  \bibfield  {author} {\bibinfo {author} {\bibfnamefont {E.}~\bibnamefont
  {Guardado-Sanchez}}, \bibinfo {author} {\bibfnamefont {P.~T.}\ \bibnamefont
  {Brown}}, \bibinfo {author} {\bibfnamefont {D.}~\bibnamefont {Mitra}},
  \bibinfo {author} {\bibfnamefont {T.}~\bibnamefont {Devakul}}, \bibinfo
  {author} {\bibfnamefont {D.~A.}\ \bibnamefont {Huse}}, \bibinfo {author}
  {\bibfnamefont {P.}~\bibnamefont {Schau{\ss}}}, \ and\ \bibinfo {author}
  {\bibfnamefont {W.~S.}\ \bibnamefont {Bakr}},\ }\href {\doibase
  10.1103/PhysRevX.8.021069} {\bibfield  {journal} {\bibinfo  {journal}
  {Physical Review X}\ }\textbf {\bibinfo {volume} {8}},\ \bibinfo {pages}
  {021069} (\bibinfo {year} {2018})}\BibitemShut {NoStop}%
\bibitem [{\citenamefont {Barredo}\ \emph {et~al.}(2018)\citenamefont
  {Barredo}, \citenamefont {Lienhard}, \citenamefont {de~L{\'{e}}s{\'{e}}leuc},
  \citenamefont {Lahaye},\ and\ \citenamefont {Browaeys}}]{Barredo2018}%
  \BibitemOpen
  \bibfield  {author} {\bibinfo {author} {\bibfnamefont {D.}~\bibnamefont
  {Barredo}}, \bibinfo {author} {\bibfnamefont {V.}~\bibnamefont {Lienhard}},
  \bibinfo {author} {\bibfnamefont {S.}~\bibnamefont
  {de~L{\'{e}}s{\'{e}}leuc}}, \bibinfo {author} {\bibfnamefont
  {T.}~\bibnamefont {Lahaye}}, \ and\ \bibinfo {author} {\bibfnamefont
  {A.}~\bibnamefont {Browaeys}},\ }\href {\doibase 10.1038/s41586-018-0450-2}
  {\bibfield  {journal} {\bibinfo  {journal} {Nature}\ }\textbf {\bibinfo
  {volume} {561}},\ \bibinfo {pages} {79} (\bibinfo {year} {2018})}\BibitemShut
  {NoStop}%
\bibitem [{SI()}]{SI}%
  \BibitemOpen
  \href@noop {} {}\bibinfo {note} {See Supplementary Information.}\BibitemShut
  {Stop}%
\bibitem [{\citenamefont {Pohl}\ \emph {et~al.}(2010)\citenamefont {Pohl},
  \citenamefont {Demler},\ and\ \citenamefont {Lukin}}]{Pohl2010}%
  \BibitemOpen
  \bibfield  {author} {\bibinfo {author} {\bibfnamefont {T.}~\bibnamefont
  {Pohl}}, \bibinfo {author} {\bibfnamefont {E.}~\bibnamefont {Demler}}, \ and\
  \bibinfo {author} {\bibfnamefont {M.~D.}\ \bibnamefont {Lukin}},\ }\href
  {\doibase 10.1103/PhysRevLett.104.043002} {\bibfield  {journal} {\bibinfo
  {journal} {Physical Review Letters}\ }\textbf {\bibinfo {volume} {104}},\
  \bibinfo {pages} {043002} (\bibinfo {year} {2010})}\BibitemShut {NoStop}%
\bibitem [{\citenamefont {van Bijnen}\ \emph {et~al.}(2011)\citenamefont {van
  Bijnen}, \citenamefont {Smit}, \citenamefont {van Leeuwen}, \citenamefont
  {Vredenbregt},\ and\ \citenamefont {Kokkelmans}}]{vanBinjen2011}%
  \BibitemOpen
  \bibfield  {author} {\bibinfo {author} {\bibfnamefont {R.~M.~W.}\
  \bibnamefont {van Bijnen}}, \bibinfo {author} {\bibfnamefont
  {S.}~\bibnamefont {Smit}}, \bibinfo {author} {\bibfnamefont {K.~A.~H.}\
  \bibnamefont {van Leeuwen}}, \bibinfo {author} {\bibfnamefont {E.~J.~D.}\
  \bibnamefont {Vredenbregt}}, \ and\ \bibinfo {author} {\bibfnamefont {S.~J.
  J. M.~F.}\ \bibnamefont {Kokkelmans}},\ }\href {\doibase
  10.1088/0953-4075/44/18/184008} {\bibfield  {journal} {\bibinfo  {journal}
  {Journal of Physics B: Atomic, Molecular and Optical Physics}\ }\textbf
  {\bibinfo {volume} {44}},\ \bibinfo {pages} {184008} (\bibinfo {year}
  {2011})}\BibitemShut {NoStop}%
\bibitem [{\citenamefont {Weimer}\ and\ \citenamefont
  {B{\"{u}}chler}(2010)}]{Weimer2010}%
  \BibitemOpen
  \bibfield  {author} {\bibinfo {author} {\bibfnamefont {H.}~\bibnamefont
  {Weimer}}\ and\ \bibinfo {author} {\bibfnamefont {H.~P.}\ \bibnamefont
  {B{\"{u}}chler}},\ }\href {\doibase 10.1103/PhysRevLett.105.230403}
  {\bibfield  {journal} {\bibinfo  {journal} {Phys. Rev. Lett.}\ }\textbf
  {\bibinfo {volume} {105}},\ \bibinfo {pages} {230403} (\bibinfo {year}
  {2010})}\BibitemShut {NoStop}%
\bibitem [{\citenamefont {Sela}\ \emph {et~al.}(2011)\citenamefont {Sela},
  \citenamefont {Punk},\ and\ \citenamefont {Garst}}]{Sela2011}%
  \BibitemOpen
  \bibfield  {author} {\bibinfo {author} {\bibfnamefont {E.}~\bibnamefont
  {Sela}}, \bibinfo {author} {\bibfnamefont {M.}~\bibnamefont {Punk}}, \ and\
  \bibinfo {author} {\bibfnamefont {M.}~\bibnamefont {Garst}},\ }\href
  {\doibase 10.1103/PhysRevB.84.085434} {\bibfield  {journal} {\bibinfo
  {journal} {Physical Review B}\ }\textbf {\bibinfo {volume} {84}},\ \bibinfo
  {pages} {085434} (\bibinfo {year} {2011})}\BibitemShut {NoStop}%
\bibitem [{\citenamefont {Bravyi}\ \emph {et~al.}(2008)\citenamefont {Bravyi},
  \citenamefont {Divincenzo}, \citenamefont {Oliveira},\ and\ \citenamefont
  {Terhal}}]{Bravyi2008}%
  \BibitemOpen
  \bibfield  {author} {\bibinfo {author} {\bibfnamefont {S.}~\bibnamefont
  {Bravyi}}, \bibinfo {author} {\bibfnamefont {D.~P.}\ \bibnamefont
  {Divincenzo}}, \bibinfo {author} {\bibfnamefont {R.}~\bibnamefont
  {Oliveira}}, \ and\ \bibinfo {author} {\bibfnamefont {B.~M.}\ \bibnamefont
  {Terhal}},\ }\href {http://dl.acm.org/citation.cfm?id=2011772.2011773}
  {\bibfield  {journal} {\bibinfo  {journal} {Quantum Info. Comput.}\ }\textbf
  {\bibinfo {volume} {8}},\ \bibinfo {pages} {361} (\bibinfo {year}
  {2008})}\BibitemShut {NoStop}%
\bibitem [{Note1()}]{Note1}%
  \BibitemOpen
  \bibinfo {note} {The exact phases required to render the Rydberg Hamiltonian
  in the form (\ref {eqn:Hryd}) will vary between different experimental
  realizations, but as the final measurements are always taken in the
  occupation number basis this has no effect on observables, provided no
  variation in the laser phase occurs during evolution.}\BibitemShut {Stop}%
\bibitem [{\citenamefont {Levine}\ \emph {et~al.}(2018)\citenamefont {Levine},
  \citenamefont {Keesling}, \citenamefont {Omran}, \citenamefont {Bernien},
  \citenamefont {Schwartz}, \citenamefont {Zibrov}, \citenamefont {Endres},
  \citenamefont {Greiner}, \citenamefont {Vuleti{\'{c}}},\ and\ \citenamefont
  {Lukin}}]{Levine2018}%
  \BibitemOpen
  \bibfield  {author} {\bibinfo {author} {\bibfnamefont {H.}~\bibnamefont
  {Levine}}, \bibinfo {author} {\bibfnamefont {A.}~\bibnamefont {Keesling}},
  \bibinfo {author} {\bibfnamefont {A.}~\bibnamefont {Omran}}, \bibinfo
  {author} {\bibfnamefont {H.}~\bibnamefont {Bernien}}, \bibinfo {author}
  {\bibfnamefont {S.}~\bibnamefont {Schwartz}}, \bibinfo {author}
  {\bibfnamefont {A.~S.}\ \bibnamefont {Zibrov}}, \bibinfo {author}
  {\bibfnamefont {M.}~\bibnamefont {Endres}}, \bibinfo {author} {\bibfnamefont
  {M.}~\bibnamefont {Greiner}}, \bibinfo {author} {\bibfnamefont
  {V.}~\bibnamefont {Vuleti{\'{c}}}}, \ and\ \bibinfo {author} {\bibfnamefont
  {M.~D.}\ \bibnamefont {Lukin}},\ }\href {\doibase
  10.1103/PhysRevLett.121.123603} {\bibfield  {journal} {\bibinfo  {journal}
  {Physical Review Letters}\ }\textbf {\bibinfo {volume} {121}},\ \bibinfo
  {pages} {123603} (\bibinfo {year} {2018})}\BibitemShut {NoStop}%
\bibitem [{\citenamefont {Ackley}\ \emph {et~al.}(1985)\citenamefont {Ackley},
  \citenamefont {Hinton},\ and\ \citenamefont {Sejnowski}}]{Ackley85}%
  \BibitemOpen
  \bibfield  {author} {\bibinfo {author} {\bibfnamefont {D.~H.}\ \bibnamefont
  {Ackley}}, \bibinfo {author} {\bibfnamefont {G.~E.}\ \bibnamefont {Hinton}},
  \ and\ \bibinfo {author} {\bibfnamefont {T.~J.}\ \bibnamefont {Sejnowski}},\
  }\href {\doibase https://doi.org/10.1016/S0364-0213(85)80012-4} {\bibfield
  {journal} {\bibinfo  {journal} {Cognitive Science}\ }\textbf {\bibinfo
  {volume} {9}},\ \bibinfo {pages} {147 } (\bibinfo {year} {1985})}\BibitemShut
  {NoStop}%
\bibitem [{\citenamefont {Smolensky}(1986)}]{Smolensky1986}%
  \BibitemOpen
  \bibfield  {author} {\bibinfo {author} {\bibfnamefont {P.}~\bibnamefont
  {Smolensky}},\ }in\ \href {http://dl.acm.org/citation.cfm?id=104279.104290}
  {\emph {\bibinfo {booktitle} {Parallel Distributed Processing}}},\ \bibinfo
  {editor} {edited by\ \bibinfo {editor} {\bibfnamefont {D.~E.}\ \bibnamefont
  {Rumelhart}}, \bibinfo {editor} {\bibfnamefont {J.~L.}\ \bibnamefont
  {McClelland}}, \ and\ \bibinfo {editor} {\bibfnamefont {C.}~\bibnamefont {PDP
  Research~Group}}}\ (\bibinfo  {publisher} {MIT Press},\ \bibinfo {address}
  {Cambridge, MA, USA},\ \bibinfo {year} {1986})\ Chap.\ \bibinfo {chapter}
  {Information Processing in Dynamical Systems: Foundations of Harmony Theory},
  pp.\ \bibinfo {pages} {194--281}\BibitemShut {NoStop}%
\bibitem [{\citenamefont {Torlai}\ and\ \citenamefont
  {Melko}(2016)}]{PhysRevB.94.165134}%
  \BibitemOpen
  \bibfield  {author} {\bibinfo {author} {\bibfnamefont {G.}~\bibnamefont
  {Torlai}}\ and\ \bibinfo {author} {\bibfnamefont {R.~G.}\ \bibnamefont
  {Melko}},\ }\href {\doibase 10.1103/PhysRevB.94.165134} {\bibfield  {journal}
  {\bibinfo  {journal} {Physical Review B}\ }\textbf {\bibinfo {volume} {94}},\
  \bibinfo {pages} {165134} (\bibinfo {year} {2016})}\BibitemShut {NoStop}%
\bibitem [{\citenamefont {Sehayek}\ \emph {et~al.}(2019)\citenamefont
  {Sehayek}, \citenamefont {Golubeva}, \citenamefont {Albergo}, \citenamefont
  {Kulchytskyy}, \citenamefont {Torlai},\ and\ \citenamefont
  {Melko}}]{Sehayek2019}%
  \BibitemOpen
  \bibfield  {author} {\bibinfo {author} {\bibfnamefont {D.}~\bibnamefont
  {Sehayek}}, \bibinfo {author} {\bibfnamefont {A.}~\bibnamefont {Golubeva}},
  \bibinfo {author} {\bibfnamefont {M.~S.}\ \bibnamefont {Albergo}}, \bibinfo
  {author} {\bibfnamefont {B.}~\bibnamefont {Kulchytskyy}}, \bibinfo {author}
  {\bibfnamefont {G.}~\bibnamefont {Torlai}}, \ and\ \bibinfo {author}
  {\bibfnamefont {R.~G.}\ \bibnamefont {Melko}},\ }\href
  {http://arxiv.org/abs/1908.07532} {\bibfield  {journal} {\bibinfo  {journal}
  {arXiv:1908.07532 [quant-ph]}\ } (\bibinfo {year} {2019})},\ \bibinfo {note}
  {arXiv: 1908.07532}\BibitemShut {NoStop}%
\bibitem [{\citenamefont {Torlai}\ \emph {et~al.}(2018)\citenamefont {Torlai},
  \citenamefont {Mazzola}, \citenamefont {Carrasquilla}, \citenamefont
  {Troyer}, \citenamefont {Melko},\ and\ \citenamefont
  {Carleo}}]{torlai_2018_nnqst}%
  \BibitemOpen
  \bibfield  {author} {\bibinfo {author} {\bibfnamefont {G.}~\bibnamefont
  {Torlai}}, \bibinfo {author} {\bibfnamefont {G.}~\bibnamefont {Mazzola}},
  \bibinfo {author} {\bibfnamefont {J.}~\bibnamefont {Carrasquilla}}, \bibinfo
  {author} {\bibfnamefont {M.}~\bibnamefont {Troyer}}, \bibinfo {author}
  {\bibfnamefont {R.}~\bibnamefont {Melko}}, \ and\ \bibinfo {author}
  {\bibfnamefont {G.}~\bibnamefont {Carleo}},\ }\href {\doibase
  10.1038/s41567-018-0048-5} {\bibfield  {journal} {\bibinfo  {journal} {Nature
  Physics}\ }\textbf {\bibinfo {volume} {14}},\ \bibinfo {pages} {447}
  (\bibinfo {year} {2018})}\BibitemShut {NoStop}%
\bibitem [{\citenamefont {Carleo}\ and\ \citenamefont
  {Troyer}(2017)}]{carleo_2017_solving_mb}%
  \BibitemOpen
  \bibfield  {author} {\bibinfo {author} {\bibfnamefont {G.}~\bibnamefont
  {Carleo}}\ and\ \bibinfo {author} {\bibfnamefont {M.}~\bibnamefont
  {Troyer}},\ }\href {\doibase 10.1126/science.aag2302} {\bibfield  {journal}
  {\bibinfo  {journal} {Science}\ }\textbf {\bibinfo {volume} {355}},\ \bibinfo
  {pages} {602} (\bibinfo {year} {2017})}\BibitemShut {NoStop}%
\bibitem [{\citenamefont {Torlai}\ and\ \citenamefont
  {Melko}(2018)}]{torlai_2018_ndo}%
  \BibitemOpen
  \bibfield  {author} {\bibinfo {author} {\bibfnamefont {G.}~\bibnamefont
  {Torlai}}\ and\ \bibinfo {author} {\bibfnamefont {R.~G.}\ \bibnamefont
  {Melko}},\ }\href {\doibase 10.1103/PhysRevLett.120.240503} {\bibfield
  {journal} {\bibinfo  {journal} {Physical Review Letters}\ }\textbf {\bibinfo
  {volume} {120}},\ \bibinfo {pages} {240503} (\bibinfo {year}
  {2018})}\BibitemShut {NoStop}%
\bibitem [{\citenamefont {Carrasquilla}\ \emph {et~al.}(2019)\citenamefont
  {Carrasquilla}, \citenamefont {Torlai}, \citenamefont {Melko},\ and\
  \citenamefont {Aolita}}]{Carrasquilla2018}%
  \BibitemOpen
  \bibfield  {author} {\bibinfo {author} {\bibfnamefont {J.}~\bibnamefont
  {Carrasquilla}}, \bibinfo {author} {\bibfnamefont {G.}~\bibnamefont
  {Torlai}}, \bibinfo {author} {\bibfnamefont {R.~G.}\ \bibnamefont {Melko}}, \
  and\ \bibinfo {author} {\bibfnamefont {L.}~\bibnamefont {Aolita}},\ }\href
  {\doibase 10.1038/s42256-019-0028-1} {\bibfield  {journal} {\bibinfo
  {journal} {Nature Machine Intelligence}\ }\textbf {\bibinfo {volume} {1}},\
  \bibinfo {pages} {155} (\bibinfo {year} {2019})}\BibitemShut {NoStop}%
\bibitem [{\citenamefont {Nielsen}\ and\ \citenamefont
  {Chuang}(2010)}]{Nielsen2011}%
  \BibitemOpen
  \bibfield  {author} {\bibinfo {author} {\bibfnamefont {M.~A.}\ \bibnamefont
  {Nielsen}}\ and\ \bibinfo {author} {\bibfnamefont {I.~L.}\ \bibnamefont
  {Chuang}},\ }\href {\doibase 10.1017/CBO9780511976667} {\emph {\bibinfo
  {title} {{Quantum Computation and Quantum Information}}}},\ \bibinfo
  {edition} {10th}\ ed.\ (\bibinfo  {publisher} {Cambridge University Press},\
  \bibinfo {address} {Cambridge},\ \bibinfo {year} {2010})\BibitemShut
  {NoStop}%
\bibitem [{\citenamefont {{Yichuan Tang}}\ \emph {et~al.}(2012)\citenamefont
  {{Yichuan Tang}}, \citenamefont {Salakhutdinov},\ and\ \citenamefont
  {Hinton}}]{YichuanTang2012}%
  \BibitemOpen
  \bibfield  {author} {\bibinfo {author} {\bibnamefont {{Yichuan Tang}}},
  \bibinfo {author} {\bibfnamefont {R.}~\bibnamefont {Salakhutdinov}}, \ and\
  \bibinfo {author} {\bibfnamefont {G.}~\bibnamefont {Hinton}},\ }in\ \href
  {\doibase 10.1109/CVPR.2012.6247936} {\emph {\bibinfo {booktitle} {2012 IEEE
  Conference on Computer Vision and Pattern Recognition}}}\ (\bibinfo
  {publisher} {IEEE},\ \bibinfo {year} {2012})\ pp.\ \bibinfo {pages}
  {2264--2271}\BibitemShut {NoStop}%
\bibitem [{\citenamefont {Islam}\ \emph {et~al.}(2015)\citenamefont {Islam},
  \citenamefont {Ma}, \citenamefont {Preiss}, \citenamefont {{Eric Tai}},
  \citenamefont {Lukin}, \citenamefont {Rispoli},\ and\ \citenamefont
  {Greiner}}]{Islam2015}%
  \BibitemOpen
  \bibfield  {author} {\bibinfo {author} {\bibfnamefont {R.}~\bibnamefont
  {Islam}}, \bibinfo {author} {\bibfnamefont {R.}~\bibnamefont {Ma}}, \bibinfo
  {author} {\bibfnamefont {P.~M.}\ \bibnamefont {Preiss}}, \bibinfo {author}
  {\bibfnamefont {M.}~\bibnamefont {{Eric Tai}}}, \bibinfo {author}
  {\bibfnamefont {A.}~\bibnamefont {Lukin}}, \bibinfo {author} {\bibfnamefont
  {M.}~\bibnamefont {Rispoli}}, \ and\ \bibinfo {author} {\bibfnamefont
  {M.}~\bibnamefont {Greiner}},\ }\href {\doibase 10.1038/nature15750}
  {\bibfield  {journal} {\bibinfo  {journal} {Nature}\ }\textbf {\bibinfo
  {volume} {528}},\ \bibinfo {pages} {77} (\bibinfo {year} {2015})}\BibitemShut
  {NoStop}%
\bibitem [{\citenamefont {Brydges}\ \emph {et~al.}(2018)\citenamefont
  {Brydges}, \citenamefont {Elben}, \citenamefont {Jurcevic}, \citenamefont
  {Vermersch}, \citenamefont {Maier}, \citenamefont {Lanyon}, \citenamefont
  {Zoller}, \citenamefont {Blatt},\ and\ \citenamefont {Roos}}]{Brydges2018}%
  \BibitemOpen
  \bibfield  {author} {\bibinfo {author} {\bibfnamefont {T.}~\bibnamefont
  {Brydges}}, \bibinfo {author} {\bibfnamefont {A.}~\bibnamefont {Elben}},
  \bibinfo {author} {\bibfnamefont {P.}~\bibnamefont {Jurcevic}}, \bibinfo
  {author} {\bibfnamefont {B.}~\bibnamefont {Vermersch}}, \bibinfo {author}
  {\bibfnamefont {C.}~\bibnamefont {Maier}}, \bibinfo {author} {\bibfnamefont
  {B.~P.}\ \bibnamefont {Lanyon}}, \bibinfo {author} {\bibfnamefont
  {P.}~\bibnamefont {Zoller}}, \bibinfo {author} {\bibfnamefont
  {R.}~\bibnamefont {Blatt}}, \ and\ \bibinfo {author} {\bibfnamefont {C.~F.}\
  \bibnamefont {Roos}},\ }\href {http://arxiv.org/abs/1806.05747} {\  (\bibinfo
  {year} {2018})},\ \Eprint {http://arxiv.org/abs/1806.05747}
  {arXiv:1806.05747} \BibitemShut {NoStop}%
\bibitem [{\citenamefont {Hastings}\ \emph {et~al.}(2010)\citenamefont
  {Hastings}, \citenamefont {Gonz{\'{a}}lez}, \citenamefont {Kallin},\ and\
  \citenamefont {Melko}}]{Hastings2010}%
  \BibitemOpen
  \bibfield  {author} {\bibinfo {author} {\bibfnamefont {M.~B.}\ \bibnamefont
  {Hastings}}, \bibinfo {author} {\bibfnamefont {I.}~\bibnamefont
  {Gonz{\'{a}}lez}}, \bibinfo {author} {\bibfnamefont {A.~B.}\ \bibnamefont
  {Kallin}}, \ and\ \bibinfo {author} {\bibfnamefont {R.~G.}\ \bibnamefont
  {Melko}},\ }\href {\doibase 10.1103/PhysRevLett.104.157201} {\bibfield
  {journal} {\bibinfo  {journal} {Physical Review Letters}\ }\textbf {\bibinfo
  {volume} {104}},\ \bibinfo {pages} {157201} (\bibinfo {year}
  {2010})}\BibitemShut {NoStop}%
\bibitem [{\citenamefont {Zhang}\ \emph {et~al.}(2011)\citenamefont {Zhang},
  \citenamefont {Grover},\ and\ \citenamefont {Vishwanath}}]{Zhang2011}%
  \BibitemOpen
  \bibfield  {author} {\bibinfo {author} {\bibfnamefont {Y.}~\bibnamefont
  {Zhang}}, \bibinfo {author} {\bibfnamefont {T.}~\bibnamefont {Grover}}, \
  and\ \bibinfo {author} {\bibfnamefont {A.}~\bibnamefont {Vishwanath}},\
  }\href {\doibase 10.1103/PhysRevLett.107.067202} {\bibfield  {journal}
  {\bibinfo  {journal} {Physical Review Letters}\ }\textbf {\bibinfo {volume}
  {107}},\ \bibinfo {pages} {067202} (\bibinfo {year} {2011})}\BibitemShut
  {NoStop}%
\bibitem [{\citenamefont {Grover}\ and\ \citenamefont
  {Fisher}(2015)}]{Grover2015}%
  \BibitemOpen
  \bibfield  {author} {\bibinfo {author} {\bibfnamefont {T.}~\bibnamefont
  {Grover}}\ and\ \bibinfo {author} {\bibfnamefont {M.~P.~A.}\ \bibnamefont
  {Fisher}},\ }\href {\doibase 10.1103/PhysRevA.92.042308} {\bibfield
  {journal} {\bibinfo  {journal} {Physical Review A}\ }\textbf {\bibinfo
  {volume} {92}},\ \bibinfo {pages} {042308} (\bibinfo {year}
  {2015})}\BibitemShut {NoStop}%
\bibitem [{\citenamefont {Bakr}\ \emph {et~al.}(2009)\citenamefont {Bakr},
  \citenamefont {Gillen}, \citenamefont {Peng}, \citenamefont {F{\"{o}}lling},\
  and\ \citenamefont {Greiner}}]{Bakr2009}%
  \BibitemOpen
  \bibfield  {author} {\bibinfo {author} {\bibfnamefont {W.~S.}\ \bibnamefont
  {Bakr}}, \bibinfo {author} {\bibfnamefont {J.~I.}\ \bibnamefont {Gillen}},
  \bibinfo {author} {\bibfnamefont {A.}~\bibnamefont {Peng}}, \bibinfo {author}
  {\bibfnamefont {S.}~\bibnamefont {F{\"{o}}lling}}, \ and\ \bibinfo {author}
  {\bibfnamefont {M.}~\bibnamefont {Greiner}},\ }\href {\doibase
  10.1038/nature08482} {\bibfield  {journal} {\bibinfo  {journal} {Nature}\
  }\textbf {\bibinfo {volume} {462}},\ \bibinfo {pages} {74} (\bibinfo {year}
  {2009})}\BibitemShut {NoStop}%
\bibitem [{\citenamefont {Weitenberg}\ \emph {et~al.}(2011)\citenamefont
  {Weitenberg}, \citenamefont {Endres}, \citenamefont {Sherson}, \citenamefont
  {Cheneau}, \citenamefont {Schau{\ss}}, \citenamefont {Fukuhara},
  \citenamefont {Bloch},\ and\ \citenamefont {Kuhr}}]{Weitenberg2011}%
  \BibitemOpen
  \bibfield  {author} {\bibinfo {author} {\bibfnamefont {C.}~\bibnamefont
  {Weitenberg}}, \bibinfo {author} {\bibfnamefont {M.}~\bibnamefont {Endres}},
  \bibinfo {author} {\bibfnamefont {J.~F.}\ \bibnamefont {Sherson}}, \bibinfo
  {author} {\bibfnamefont {M.}~\bibnamefont {Cheneau}}, \bibinfo {author}
  {\bibfnamefont {P.}~\bibnamefont {Schau{\ss}}}, \bibinfo {author}
  {\bibfnamefont {T.}~\bibnamefont {Fukuhara}}, \bibinfo {author}
  {\bibfnamefont {I.}~\bibnamefont {Bloch}}, \ and\ \bibinfo {author}
  {\bibfnamefont {S.}~\bibnamefont {Kuhr}},\ }\href {\doibase
  10.1038/nature09827} {\bibfield  {journal} {\bibinfo  {journal} {Nature}\
  }\textbf {\bibinfo {volume} {471}},\ \bibinfo {pages} {319} (\bibinfo {year}
  {2011})}\BibitemShut {NoStop}%
\bibitem [{\citenamefont {Kaufman}\ \emph {et~al.}(2016)\citenamefont
  {Kaufman}, \citenamefont {Tai}, \citenamefont {Lukin}, \citenamefont
  {Rispoli}, \citenamefont {Schittko}, \citenamefont {Preiss},\ and\
  \citenamefont {Greiner}}]{Kaufman2016}%
  \BibitemOpen
  \bibfield  {author} {\bibinfo {author} {\bibfnamefont {A.~M.}\ \bibnamefont
  {Kaufman}}, \bibinfo {author} {\bibfnamefont {M.~E.}\ \bibnamefont {Tai}},
  \bibinfo {author} {\bibfnamefont {A.}~\bibnamefont {Lukin}}, \bibinfo
  {author} {\bibfnamefont {M.}~\bibnamefont {Rispoli}}, \bibinfo {author}
  {\bibfnamefont {R.}~\bibnamefont {Schittko}}, \bibinfo {author}
  {\bibfnamefont {P.~M.}\ \bibnamefont {Preiss}}, \ and\ \bibinfo {author}
  {\bibfnamefont {M.}~\bibnamefont {Greiner}},\ }\href {\doibase
  10.1126/science.aaf6725} {\bibfield  {journal} {\bibinfo  {journal}
  {Science}\ }\textbf {\bibinfo {volume} {353}},\ \bibinfo {pages} {794}
  (\bibinfo {year} {2016})}\BibitemShut {NoStop}%
\bibitem [{\citenamefont {Cheuk}\ \emph {et~al.}(2015)\citenamefont {Cheuk},
  \citenamefont {Nichols}, \citenamefont {Okan}, \citenamefont {Gersdorf},
  \citenamefont {Ramasesh}, \citenamefont {Bakr}, \citenamefont {Lompe},\ and\
  \citenamefont {Zwierlein}}]{Cheuk2015}%
  \BibitemOpen
  \bibfield  {author} {\bibinfo {author} {\bibfnamefont {L.~W.}\ \bibnamefont
  {Cheuk}}, \bibinfo {author} {\bibfnamefont {M.~A.}\ \bibnamefont {Nichols}},
  \bibinfo {author} {\bibfnamefont {M.}~\bibnamefont {Okan}}, \bibinfo {author}
  {\bibfnamefont {T.}~\bibnamefont {Gersdorf}}, \bibinfo {author}
  {\bibfnamefont {V.~V.}\ \bibnamefont {Ramasesh}}, \bibinfo {author}
  {\bibfnamefont {W.~S.}\ \bibnamefont {Bakr}}, \bibinfo {author}
  {\bibfnamefont {T.}~\bibnamefont {Lompe}}, \ and\ \bibinfo {author}
  {\bibfnamefont {M.~W.}\ \bibnamefont {Zwierlein}},\ }\href {\doibase
  10.1103/PhysRevLett.114.193001} {\bibfield  {journal} {\bibinfo  {journal}
  {Physical Review Letters}\ }\textbf {\bibinfo {volume} {114}},\ \bibinfo
  {pages} {193001} (\bibinfo {year} {2015})}\BibitemShut {NoStop}%
\bibitem [{\citenamefont {Greif}\ \emph {et~al.}(2016)\citenamefont {Greif},
  \citenamefont {Parsons}, \citenamefont {Mazurenko}, \citenamefont {Chiu},
  \citenamefont {Blatt}, \citenamefont {Huber}, \citenamefont {Ji},\ and\
  \citenamefont {Greiner}}]{Greif2016}%
  \BibitemOpen
  \bibfield  {author} {\bibinfo {author} {\bibfnamefont {D.}~\bibnamefont
  {Greif}}, \bibinfo {author} {\bibfnamefont {M.~F.}\ \bibnamefont {Parsons}},
  \bibinfo {author} {\bibfnamefont {A.}~\bibnamefont {Mazurenko}}, \bibinfo
  {author} {\bibfnamefont {C.~S.}\ \bibnamefont {Chiu}}, \bibinfo {author}
  {\bibfnamefont {S.}~\bibnamefont {Blatt}}, \bibinfo {author} {\bibfnamefont
  {F.}~\bibnamefont {Huber}}, \bibinfo {author} {\bibfnamefont
  {G.}~\bibnamefont {Ji}}, \ and\ \bibinfo {author} {\bibfnamefont
  {M.}~\bibnamefont {Greiner}},\ }\href {\doibase 10.1126/science.aad9041}
  {\bibfield  {journal} {\bibinfo  {journal} {Science (New York, N.Y.)}\
  }\textbf {\bibinfo {volume} {351}},\ \bibinfo {pages} {953} (\bibinfo {year}
  {2016})}\BibitemShut {NoStop}%
\bibitem [{\citenamefont {Sachdev}(2011)}]{Sachdev2011}%
  \BibitemOpen
  \bibfield  {author} {\bibinfo {author} {\bibfnamefont {S.}~\bibnamefont
  {Sachdev}},\ }\href {\doibase 10.1017/CBO9780511973765} {\emph {\bibinfo
  {title} {Quantum Phase Transitions}}},\ \bibinfo {edition} {2nd}\ ed.\
  (\bibinfo  {publisher} {Cambridge University Press},\ \bibinfo {year}
  {2011})\BibitemShut {NoStop}%
\bibitem [{\citenamefont {Hinton}(2012)}]{Hinton2012}%
  \BibitemOpen
  \bibfield  {author} {\bibinfo {author} {\bibfnamefont {G.~E.}\ \bibnamefont
  {Hinton}},\ }in\ \href {\doibase 10.1007/978-3-642-35289-8_32} {\emph
  {\bibinfo {booktitle} {Neural Networks: Tricks of the Trade: Second
  Edition}}},\ \bibinfo {editor} {edited by\ \bibinfo {editor} {\bibfnamefont
  {G.}~\bibnamefont {Montavon}}, \bibinfo {editor} {\bibfnamefont {G.~B.}\
  \bibnamefont {Orr}}, \ and\ \bibinfo {editor} {\bibfnamefont {K.-R.}\
  \bibnamefont {M{\"{u}}ller}}}\ (\bibinfo  {publisher} {Springer Berlin
  Heidelberg},\ \bibinfo {address} {Berlin, Heidelberg},\ \bibinfo {year}
  {2012})\ pp.\ \bibinfo {pages} {599--619}\BibitemShut {NoStop}%
\bibitem [{\citenamefont {Hinton}(2002)}]{Hinton2002}%
  \BibitemOpen
  \bibfield  {author} {\bibinfo {author} {\bibfnamefont {G.~E.}\ \bibnamefont
  {Hinton}},\ }\href {\doibase 10.1162/089976602760128018} {\bibfield
  {journal} {\bibinfo  {journal} {Neural Comput.}\ }\textbf {\bibinfo {volume}
  {14}},\ \bibinfo {pages} {1771} (\bibinfo {year} {2002})}\BibitemShut
  {NoStop}%
\bibitem [{\citenamefont {Beach}\ \emph {et~al.}(2019)\citenamefont {Beach},
  \citenamefont {Vlugt}, \citenamefont {Golubeva}, \citenamefont {Huembeli},
  \citenamefont {Kulchytskyy}, \citenamefont {Luo}, \citenamefont {Melko},
  \citenamefont {Merali},\ and\ \citenamefont {Torlai}}]{qucumber}%
  \BibitemOpen
  \bibfield  {author} {\bibinfo {author} {\bibfnamefont {M.~J.~S.}\
  \bibnamefont {Beach}}, \bibinfo {author} {\bibfnamefont {I.~D.}\ \bibnamefont
  {Vlugt}}, \bibinfo {author} {\bibfnamefont {A.}~\bibnamefont {Golubeva}},
  \bibinfo {author} {\bibfnamefont {P.}~\bibnamefont {Huembeli}}, \bibinfo
  {author} {\bibfnamefont {B.}~\bibnamefont {Kulchytskyy}}, \bibinfo {author}
  {\bibfnamefont {X.}~\bibnamefont {Luo}}, \bibinfo {author} {\bibfnamefont
  {R.~G.}\ \bibnamefont {Melko}}, \bibinfo {author} {\bibfnamefont
  {E.}~\bibnamefont {Merali}}, \ and\ \bibinfo {author} {\bibfnamefont
  {G.}~\bibnamefont {Torlai}},\ }\href {\doibase 10.21468/SciPostPhys.7.1.009}
  {\bibfield  {journal} {\bibinfo  {journal} {SciPost Phys.}\ }\textbf
  {\bibinfo {volume} {7}},\ \bibinfo {pages} {9} (\bibinfo {year}
  {2019})}\BibitemShut {NoStop}%
\bibitem [{\citenamefont {Hinton}\ \emph {et~al.}(2006)\citenamefont {Hinton},
  \citenamefont {Osindero},\ and\ \citenamefont {Teh}}]{Hinton2006DBN}%
  \BibitemOpen
  \bibfield  {author} {\bibinfo {author} {\bibfnamefont {G.~E.}\ \bibnamefont
  {Hinton}}, \bibinfo {author} {\bibfnamefont {S.}~\bibnamefont {Osindero}}, \
  and\ \bibinfo {author} {\bibfnamefont {Y.-W.}\ \bibnamefont {Teh}},\ }\href
  {\doibase 10.1162/neco.2006.18.7.1527} {\bibfield  {journal} {\bibinfo
  {journal} {Neural Computation}\ }\textbf {\bibinfo {volume} {18}},\ \bibinfo
  {pages} {1527} (\bibinfo {year} {2006})}\BibitemShut {NoStop}%
\bibitem [{\citenamefont {{Le Roux}}\ and\ \citenamefont
  {Bengio}(2008)}]{LeRoux2008}%
  \BibitemOpen
  \bibfield  {author} {\bibinfo {author} {\bibfnamefont {N.}~\bibnamefont {{Le
  Roux}}}\ and\ \bibinfo {author} {\bibfnamefont {Y.}~\bibnamefont {Bengio}},\
  }\href {\doibase 10.1162/neco.2008.04-07-510} {\bibfield  {journal} {\bibinfo
   {journal} {Neural Computation}\ }\textbf {\bibinfo {volume} {20}},\ \bibinfo
  {pages} {1631} (\bibinfo {year} {2008})}\BibitemShut {NoStop}%
\bibitem [{\citenamefont {Deng}\ \emph {et~al.}(2017)\citenamefont {Deng},
  \citenamefont {Li},\ and\ \citenamefont {{Das
  Sarma}}}]{deng_2017_rbm_entanglement}%
  \BibitemOpen
  \bibfield  {author} {\bibinfo {author} {\bibfnamefont {D.-L.}\ \bibnamefont
  {Deng}}, \bibinfo {author} {\bibfnamefont {X.}~\bibnamefont {Li}}, \ and\
  \bibinfo {author} {\bibfnamefont {S.}~\bibnamefont {{Das Sarma}}},\ }\href
  {\doibase 10.1103/PhysRevX.7.021021} {\bibfield  {journal} {\bibinfo
  {journal} {Physical Review X}\ }\textbf {\bibinfo {volume} {7}},\ \bibinfo
  {pages} {021021} (\bibinfo {year} {2017})}\BibitemShut {NoStop}%
\bibitem [{\citenamefont {Glasser}\ \emph {et~al.}(2018)\citenamefont
  {Glasser}, \citenamefont {Pancotti}, \citenamefont {August}, \citenamefont
  {Rodriguez},\ and\ \citenamefont {Cirac}}]{Glasser2018}%
  \BibitemOpen
  \bibfield  {author} {\bibinfo {author} {\bibfnamefont {I.}~\bibnamefont
  {Glasser}}, \bibinfo {author} {\bibfnamefont {N.}~\bibnamefont {Pancotti}},
  \bibinfo {author} {\bibfnamefont {M.}~\bibnamefont {August}}, \bibinfo
  {author} {\bibfnamefont {I.~D.}\ \bibnamefont {Rodriguez}}, \ and\ \bibinfo
  {author} {\bibfnamefont {J.~I.}\ \bibnamefont {Cirac}},\ }\href {\doibase
  10.1103/PhysRevX.8.011006} {\bibfield  {journal} {\bibinfo  {journal}
  {Physical Review X}\ }\textbf {\bibinfo {volume} {8}},\ \bibinfo {pages}
  {011006} (\bibinfo {year} {2018})}\BibitemShut {NoStop}%
\bibitem [{\citenamefont {Chen}\ \emph {et~al.}(2018)\citenamefont {Chen},
  \citenamefont {Cheng}, \citenamefont {Xie}, \citenamefont {Wang},\ and\
  \citenamefont {Xiang}}]{Chen2018}%
  \BibitemOpen
  \bibfield  {author} {\bibinfo {author} {\bibfnamefont {J.}~\bibnamefont
  {Chen}}, \bibinfo {author} {\bibfnamefont {S.}~\bibnamefont {Cheng}},
  \bibinfo {author} {\bibfnamefont {H.}~\bibnamefont {Xie}}, \bibinfo {author}
  {\bibfnamefont {L.}~\bibnamefont {Wang}}, \ and\ \bibinfo {author}
  {\bibfnamefont {T.}~\bibnamefont {Xiang}},\ }\href {\doibase
  10.1103/PhysRevB.97.085104} {\bibfield  {journal} {\bibinfo  {journal}
  {Physical Review B}\ }\textbf {\bibinfo {volume} {97}},\ \bibinfo {pages}
  {085104} (\bibinfo {year} {2018})}\BibitemShut {NoStop}%
\bibitem [{\citenamefont {Weinberg}\ and\ \citenamefont
  {Bukov}(2017)}]{Weinberg2017}%
  \BibitemOpen
  \bibfield  {author} {\bibinfo {author} {\bibfnamefont {P.}~\bibnamefont
  {Weinberg}}\ and\ \bibinfo {author} {\bibfnamefont {M.}~\bibnamefont
  {Bukov}},\ }\href {\doibase 10.21468/SciPostPhys.2.1.003} {\bibfield
  {journal} {\bibinfo  {journal} {SciPost Physics}\ }\textbf {\bibinfo {volume}
  {2}},\ \bibinfo {pages} {003} (\bibinfo {year} {2017})}\BibitemShut {NoStop}%
\bibitem [{\citenamefont {Greig}\ \emph {et~al.}(1989)\citenamefont {Greig},
  \citenamefont {Porteous},\ and\ \citenamefont {Seheult}}]{Greig1989}%
  \BibitemOpen
  \bibfield  {author} {\bibinfo {author} {\bibfnamefont {D.~M.}\ \bibnamefont
  {Greig}}, \bibinfo {author} {\bibfnamefont {B.~T.}\ \bibnamefont {Porteous}},
  \ and\ \bibinfo {author} {\bibfnamefont {A.~H.}\ \bibnamefont {Seheult}},\
  }\href {\doibase 10.1111/j.2517-6161.1989.tb01764.x} {\bibfield  {journal}
  {\bibinfo  {journal} {Journal of the Royal Statistical Society: Series B
  (Methodological)}\ }\textbf {\bibinfo {volume} {51}},\ \bibinfo {pages} {271}
  (\bibinfo {year} {1989})}\BibitemShut {NoStop}%
\bibitem [{\citenamefont {Geman}\ and\ \citenamefont
  {Geman}(1984)}]{Geman1984}%
  \BibitemOpen
  \bibfield  {author} {\bibinfo {author} {\bibfnamefont {S.}~\bibnamefont
  {Geman}}\ and\ \bibinfo {author} {\bibfnamefont {D.}~\bibnamefont {Geman}},\
  }\href {\doibase 10.1109/TPAMI.1984.4767596} {\bibfield  {journal} {\bibinfo
  {journal} {IEEE Transactions on Pattern Analysis and Machine Intelligence}\
  }\textbf {\bibinfo {volume} {PAMI-6}},\ \bibinfo {pages} {721} (\bibinfo
  {year} {1984})}\BibitemShut {NoStop}%
\bibitem [{\citenamefont {Batson}\ and\ \citenamefont
  {Royer}(2019)}]{Batson2019}%
  \BibitemOpen
  \bibfield  {author} {\bibinfo {author} {\bibfnamefont {J.}~\bibnamefont
  {Batson}}\ and\ \bibinfo {author} {\bibfnamefont {L.}~\bibnamefont {Royer}},\
  }\href {http://arxiv.org/abs/1901.11365} {\  (\bibinfo {year} {2019})},\
  \Eprint {http://arxiv.org/abs/1901.11365} {arXiv:1901.11365} \BibitemShut
  {NoStop}%
\bibitem [{\citenamefont {Johansson}\ \emph {et~al.}(2013)\citenamefont
  {Johansson}, \citenamefont {Nation},\ and\ \citenamefont {Nori}}]{qutip}%
  \BibitemOpen
  \bibfield  {author} {\bibinfo {author} {\bibfnamefont {J.~R.}\ \bibnamefont
  {Johansson}}, \bibinfo {author} {\bibfnamefont {P.~D.}\ \bibnamefont
  {Nation}}, \ and\ \bibinfo {author} {\bibfnamefont {F.}~\bibnamefont
  {Nori}},\ }\href {\doibase https://doi.org/10.1016/j.cpc.2012.11.019}
  {\bibfield  {journal} {\bibinfo  {journal} {Computer Physics Communications}\
  }\textbf {\bibinfo {volume} {184}},\ \bibinfo {pages} {1234} (\bibinfo {year}
  {2013})}\BibitemShut {NoStop}%
\end{thebibliography}%
\vspace{1cm}

\begin{center}
\textbf{\large Supplemental Information}
\end{center}

In this Supplementary Information, we first provide a derivation of the approximate eight-atom ordered ground state. Next, we discuss how the unsupervised RBM learning process is carried out on experimental datasets, and demonstrate how the networks generalize from the finite datasets used in training. We also detail a regularization method used to mitigate the effect of measurement errors in the training set and provide numerical evidence that this technique significantly improves the fidelity of state reconstruction from noisy data. Finally, we examine how intrinsic decoherence processes impact the quality of the pure-state reconstruction procedure. An appendix provides proofs of two bounds regarding the fidelity and entanglement properties of reconstructions.

\section{Approximate Eight-atom ground state}

The full Rydberg Hamiltonian is
\begin{align}
\hat H(\Omega, \Delta) = -\Delta \sum_{i=1}^N \hat n_i - \frac{\Omega}{2} \sum_{i=1}^N \hat \sigma_i^x + \sum_{i<j} \frac{V_{nn}}{|i-j|^6} \hat n_i \hat n_j\label{eqn:Hryd}
\end{align}
At the end of the experimental sweep, the Hamiltonian has a positive detuning and a small transverse field: $\Delta > 0, V_{nn} \gg \Delta \gg |\Omega|$; furthermore, interactions between sites separated by more than two lattice spacings may be neglected, as they are weak compared to the frequencies which characterize the sweep profile. In this regime the four-excitation states
\begin{align}
\ket{e_1} &= \ket{r\:g\:r\:g\:g\:r\:g\:r} \\
\ket{e_2} &= \ket{r\:g\:g\:r\:g\:r\:g\:r} \\
\ket{e_3} &= \ket{r\:g\:r\:g\:r\:g\:g\:r}
\end{align}
 are degenerate under the classical part of the Hamiltonian $-\Delta \sum_{i=1}^N \hat n_i  + \sum_{i<j} \frac{V_{nn}}{|i-j|^6} \hat n_i \hat n_j$. The ground state lies in the subspace spanned by these three states: adding or removing an excitation requires an energy penalty proportional to $V_{nn}$ or $\Delta$ respectively. This degeneracy is lifted by a nonzero transverse field, which couples the blockaded states at second order in $\Omega$ through the three-excitation subspace. Using perturbation theory, an effective Hamiltonian~\cite{Sachdev2011} $H_{\textnormal{eff}}$ may be constructed for the blockaded subspace, whose nonzero matrix elements are given by
\begin{align}
\bra{e_1} H_{\textnormal{eff}} \ket{e_2} = \bra{e_1} H_{\textnormal{eff}} \ket{e_3} &= - \frac{\Omega^2}{4\Delta} 
\end{align}
The corresponding ground state is $\ket{\Psi} = \frac{1}{\sqrt{2}} \ket{e_1} + \frac{1}{2} \left( \ket{e_2} + \ket{e_3} \right)$.

\section{Reconstruction Methods}

\subsection{Note on terminology}
	Below we discuss strategies for training on experimental data which has been corrupted by a fixed, known noise process. $\bsig$ will denote the variables prior to corruption by measurement errors, while $\btau$ will denote those which have been subjected to the noise channel -- that is, for a fixed true value $\bsig$, the noisy outputs are distributed according to $p(\btau | \bsig)$.
	In our experiment, $\btau$ are the only accessible variables, which yield the bitstrings recorded in each dataset. A model with parameters $\bm{\lambda}$ specifies a distribution $\pt(\bsig)$ over the uncorrupted variables $\bsig$, and a corresponding corrupted distribution $\pc(\btau) = \sum_{\bsig} p(\btau | \bsig) \pt(\bsig)$.
	
\label{app:training}
\subsection{Standard RBM training method}

The standard training method involves fitting the RBM distribution $\pt (\bsig) = \frac{1}{Z_{\bm{\lambda}}} \sum_{\bh}  e^{\:\bm{h}^\top \bm{W} \bm{\sigma} + \bm{b} \cdot \bm{\sigma} + \bm{c} \cdot \bh}$ directly to the experimental datasets; in other words, it assumes a noise-free source of data:
\begin{align}
	p(\btau | \bsig) = \delta_{\btau, \bsig}
	\end{align}
	 The optimal parameters $\bm{\lambda} = \{ \mathbf{W}, \mathbf{b}, \mathbf{c} \}$ for which the RBM best reproduces the measurement data are found by minimizing the negative log-likelihood 
\begin{align}
			\mathcal{L}_{\bm{\lambda}} = - \frac{1}{|\mathcal{D}|} \sum_{\btau \in \mathcal{D}} \log \pt (\btau)
			\end{align}
			 of the RBM distribution $\pt$ averaged over the dataset $\mathcal{D}$ ($|\mathcal{D}|$ denotes the size of the dataset).
The gradient of the log-likelihood cost function with respect to the trainable parameters $\bm{\lambda}$ may be written
\begin{align}
			    \nabla_{\bm \lambda} \mathcal{L}_{\bm{\lambda}}  &= \langle  \nabla_{\bm \lambda} \mathcal{E}_{\textnormal{eff}}(\bsig) \rangle_{\pt(\bsig)} - \frac{1}{|\mathcal{D}|} \sum_{\btau \in D}  \nabla_{\bm \lambda}  \mathcal{E}_{\textnormal{eff}}(\btau) 
			    \label{eqn:two_layer_cost_gradient}
			    \end{align}
where $\langle \cdot \rangle_{\pt (\bsig)}$ denotes the expectation value with respect to the distribution $\pt (\bsig)$, and the effective energies
\begin{align}
	\mathcal{E}_{\textnormal{eff}}(\bsig) = \bb \cdot \bsig + \sum_j \log \left( 1 + e^{W_{ji} \sigma_i + c_j} \right)
\end{align}
are defined by $\pt (\bsig) = \frac{1}{Z_{\blda}} e^{\mathcal{E}_{\textnormal{eff}}(\bsig)}$. 

The second term in the cost function gradient (\ref{eqn:two_layer_cost_gradient}) is estimated using a batch of samples $\btau_i$ of size $M$ drawn from the training set $\mathcal{D}$:
\begin{align}
 \frac{1}{|\mathcal{D}|} \sum_{\btau \in D}  \nabla_{\bm \lambda}  \mathcal{E}_{\textnormal{eff}}(\btau) \approx  \frac{1}{M} \sum_{i=1}^M  \nabla_{\bm \lambda}  \mathcal{E}_{\textnormal{eff}}(\btau_i) 
\end{align}
Exact computation of the expectation value with respect to $\pt(\bsig)$ requires summing over a number of configurations which is exponential in the system size, and is therefore not tractable.
It can also be approximated by drawing $M$ samples $\bsig_i$ distributed according to $\pt(\bsig)$ and using the estimator
\begin{align}
\langle  \nabla_{\bm \lambda} \mathcal{E}_{\textnormal{eff}}(\bsig) \rangle_{\pt(\bsig)} \approx \frac{1}{M} \sum_{i=1}^M   \nabla_{\bm \lambda} \mathcal{E}_{\textnormal{eff}}(\bsig_i)
\end{align}
In principle, samples which obey the model distribution $\pt(\bsig)$ can be generated by block Gibbs sampling~\cite{Hinton2012}, which involves repeatedly sampling from the conditional distributions $\pt (\bsig | \bh)$ and $\pt (\bh | \bsig)$. Because of the \textit{restricted} nature of the RBM graph -- there are no intra-layer connections -- the conditional distributions factorize and each unit in a given layer can be exactly sampled simultaneously. In pseudocode, starting from a `seed' visible state $\bsig_1$, the Gibbs sampling algorithm is:
\begin{algorithmic}
\For{ $i$ in $[1, 2, ..., k]$}
\State Sample $\bh_i$  from $\pt(\bh| \bsig_i)$
\State Sample $\bsig_{i+1}$ from $\pt(\bsig | \bh_i)$
\EndFor
\end{algorithmic}
-- the output of the algorithm is the visible state $\bsig_{k+1}$, which will obey the model distribution $\pt(\bsig)$ for a sufficiently large number of sampling steps. In practice, the contrastive divergence algorithm~\cite{Hinton2002} is applied, where the visible state is seeded with samples from the training set, and only a small number of sampling steps $k$ is used. In practice, moderate values $k \sim 10$ are sufficient for training with stochastic gradient descent. Additional information about the RBM and its training can be found in Ref.~\cite{rbm_nisq}. An open-source software library for RBM reconstruction of generic wavefunctions is also available~\cite{qucumber}.

\subsection{Noise-regularized training method}
	In the case where the training set is known to be corrupted by a noise process $p(\btau | \bsig)$, our goal is to learn a model $\pt(\bsig)$ whose corresponding \textit{noise-corrupted} distribution $\pc (\btau)$ fits the observed data. We therefore define the corresponding log-likelihood cost function 
	\begin{align}
			\mathcal{L}_{\bm{\lambda}} = - \frac{1}{|\mathcal{D}|} \sum_{\btau \in \mathcal{D}} \log \pc (\btau)
			\end{align}
	and train the network to minimize it on each dataset. The cost gradient takes a form nearly identical to that of the standard training method (\ref{eqn:two_layer_cost_gradient}), 
		  \begin{align}
			    \nabla_{\bm \lambda} \mathcal{L}_{\bm{\lambda}}  &= \langle  \nabla_{\bm \lambda} \mathcal{E}_{\textnormal{eff}}(\bsig) \rangle_{\pt (\bsig)} - \frac{1}{|\mathcal{D}|} \sum_{\btau \in D} \langle \nabla_{\bm \lambda}  \mathcal{E}_{\textnormal{eff}}(\bsig) \rangle_{\pc (\bsig| \btau)}
			    \label{eqn:three_layer_cost_gradient}
			    \end{align}
	The second term in the gradient update step is now computed not directly from the training set samples $\btau \in \mathcal{D}$, but rather from the Bayesian posterior distribution
	\begin{align}	
		\pc (\bsig | \btau) = \frac{ p(\btau | \bsig) \pt (\bsig)}{ \pc (\btau)}
		\end{align}
		which the RBM assigns to visible states $\bsig$, given an observation $\btau$ in the noisy training set. 
		
		This alteration to the cost gradient may be viewed as a regularization of the training based on prior knowledge of the sampling process. Regularization in machine learning generally refers to techniques for improving the generalization performance of a model trained on a particular data set to new datasets drawn from the `ground truth' source. A typical regularization scheme like weight decay does not specify \textit{a priori} how the in-sample and true distributions differ, and therefore typically requires some sort of validation process -- testing the model on held-out data -- to select good hyperparameters. In contrast, our regularization method is applied in a context where all accessible datasets are corrupted by the same noise process. This makes validation as a means of selecting regularization hyperparameters impossible -- but if the noise process is known, this is no obstacle as there are no free hyperparameters to select. 
		
	In applying equation (\ref{eqn:three_layer_cost_gradient}) to the unsupervised training of an RBM, both contributions to the gradient now require computation of expectation values over marginalized distributions $\pt (\bsig)$, $\pc (\bsig | \btau)$ of the RBM, and are therefore intractable to compute exactly.	As in the noise-free training case, this problem may be circumvented using the contrastive divergence method: the first term is estimated by repeated sampling from the conditional distributions $\pt (\bsig | \bh), \pt (\bh | \bsig)$, while the second uses the same alternating sampling from the `data-clamped' distributions $\pc (\bsig | \bh, \btau), \pc (\bh | \bsig, \btau) = \pt (\bh | \bsig)$. As noted above, $\pt (\bsig | \bh), \pt(\bh | \bsig)$ are both efficiently computable due to the restricted structure of the RBM layers $\bsig, \bh$. Similarly, $\pc (\bsig | \bh, \btau)$ is efficiently computable if the error probabilities satisfy a weaker condition, namely factorizing over the uncorrupted variables:
	\begin{align}
		p(\btau | \bsig) = \prod_i p(\btau | \sigma_i)
		\label{eqn:error_factorization}
	\end{align}
	In this case, the clamped distribution may be computed explicitly as 
	\begin{align}
		\pc (\bsig | \btau, \bh) &= \prod_i \frac{ p(\btau | \sigma_i) \pt (\sigma_i | \bh)}{\sum_{\sigma_i' = 0, 1} p(\btau | \sigma_i') \pt (\sigma_i' | \bh) }  \\
		&= \prod_i \pc (\sigma_i | \btau, \bh), 
	\end{align}
amenable to efficient block-Gibbs sampling.

	\begin{figure}[t]
		\includegraphics[width=\columnwidth]{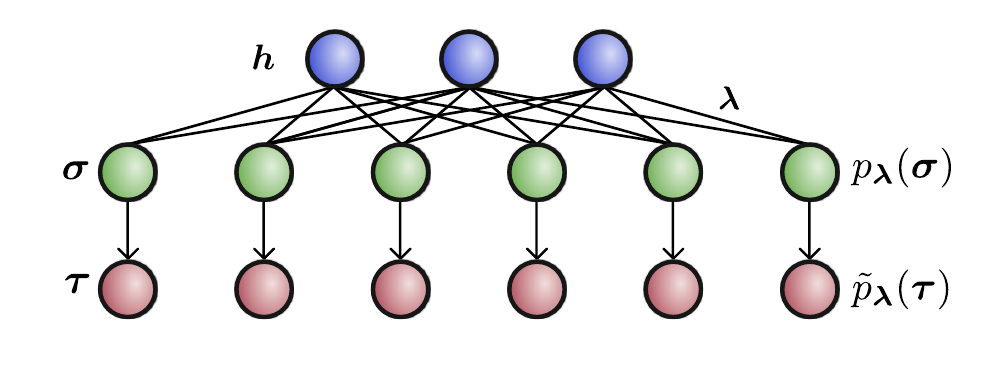}
		\caption{\textbf{Three layer model}.  Schematic for how noise-corrupted data is modeled using a three-layer graph. The upper two layers $\bh, \bsig$ constitute an RBM with trainable parameters $\blda$, which defines a distribution $\pt (\bsig)$ over the uncorrupted variables $\bsig$ upon tracing out the hidden units $\bh$. The corrupted distribution is obtained through the noise process $p(\btau | \bsig)$ as $\pc (\btau)$.   The noise process is indicated here by arrows which link uncorrupted and corrupted variables at each site. } 
		\label{fig:supp:threelayer_schematic}
	\end{figure}
	
Fig.~\ref{fig:supp:threelayer_schematic} provides an intuitive way to understand the noise regularization -- the corrupted variables $\btau$ may be included as a third \textit{noise layer} appended to the standard, two-layer RBM graph, with conditional probabilities depending on the $\bsig$ layer only. These can be interpreted as effective biases for the noise layer, which depend on the uncorrupted variables -- for example, the independent bit-flip errors used to model our Rydberg experiment may be written as
	    \begin{align*}
	    	p(\btau | \bsig) &= \frac{1}{\tilde Z} e^{\tilde{b}_\sigma \cdot \bsig + \tilde{b}_{\btau} \cdot \btau + \tilde{W} \bsig \cdot \btau } \\
	    	\tilde W &= \log \frac{p(1|1) p(0|0)}{p(1|0) p(0|1)} \\
	    	\tilde b_{\sigma,i} &= \log \frac{p(0|1)}{p(0|0)} \\
    	 	\tilde b_{\tau, i} &= \log \frac{p(1|0)}{p(0|0)} 
	    \end{align*}
	    
For brevity, we will sometimes refer to RBMs trained with this regularization as `three-layer' machines, as opposed to their `two-layer' counterparts trained in the standard fashion.  Similar graphical models known as Deep Belief Nets~\cite{Hinton2006DBN} have previously been used for unsupervised learning tasks, but with a different, layer-wise training algorithm that does not incorporate prior information; a gated RBM architecture similar to the three-layer machine has also been applied to Gaussian noise models in occluded images~\cite{YichuanTang2012}.
	    
\subsection{Sampling from trained RBMs}

After an RBM has been trained, new configurations of the uncorrupted variables $\{\bm{\sigma}\}$ can be drawn from the distribution $\pt (\bm{\sigma})$ using the block-Gibbs sampling techniques discussed above. The expectation value of a generic observable $\hat{\mathcal{O}}$ in the state $\psi_{\bm{\lambda}}(\bm{\sigma})= \sqrt{\pt (\bm{\sigma})}$ can then be approximated with a Monte Carlo average over $n_{\mathrm{mc}}$ samples:
\begin{align}
\langle \hat{\mathcal{O}} \rangle_{\psi_{\bm{\lambda}}} &= \sum_{\bsig, \bsig'} \psi_{\bm{\lambda}}(\bsig) \bra{\bsig} \hat{\mathcal{O}} \ket{\bsig'}  \psi_{\bm{\lambda}}(\bsig') \\
&= \sum_{\bsig} |\psi_{\bm{\lambda}}(\bsig)|^2 \sum_{\bsig'} \bra{\bsig} \hat{\mathcal{O}} \ket{\bsig'} \frac{ \psi_{\bm{\lambda}}(\bsig')}{\psi_{\bm{\lambda}}(\bsig)} \\
&:= \langle \mathcal{O}_L(\bm{\sigma}) \rangle_{\pt(\bsig)} \\
&\simeq n_{\mathrm{mc}}^{-1}\sum_{k=1}^{n_{\mathrm{mc}}}\mathcal{O}_L(\bm{\sigma}_k)
\end{align}
where the ``local estimate'' of the observable is defined to be
$\mathcal{O}_L(\bm{\sigma})=\sum_{\bm{\sigma}^\prime} \langle\bm{\sigma}|\hat{\mathcal{O}}|\bm{\sigma}^\prime\rangle \frac{\psi_{\bm{\lambda}}(\bm{\sigma}^\prime)}{\psi_{\bm{\lambda}}(\bm{\sigma})}$. In the case of nontrivial noise processes, to sample from the corrupted distribution $\pc (\btau)$ one may first generate an uncorrupted batch $\{\bsig \}$ of data and then sample once from the conditional distribution $p(\btau | \bsig)$ for each uncorrupted configuration.

For an RBM with $N$ visible and $N_h$ hidden units, the times for training and Monte Carlo observable estimation scale as $\mathcal{O}(N N_h)$, or in terms of the model complexity $\alpha =N_h / N$, as $\mathcal{O}( \alpha N^2)$; note that the number of visible units is fixed by the system size. A universal approximation theorem~\cite{LeRoux2008} guarantees that RBMs can represent any distribution over binary variables, although an exponentially large number of hidden units may be required in general.
However, many quantum states relevant to experiment, such as ground states of paradigmatic Hamiltonians and some matrix product states, have been found to admit efficient descriptions~\cite{torlai_2018_nnqst,PhysRevB.94.165134,deng_2017_rbm_entanglement,Glasser2018,Chen2018}. In the present work with eight atoms, the Hilbert space is small enough that
all amplitudes and expectation values can be
computed exactly, providing a valuable check on our procedure.
Such a benchmark quickly becomes impossible with current classical hardware when the number of atoms approaches $\sim 20$ for pure states,
and at even smaller chain lengths for the exact evaluation of non-pure states.

\section{Training details}
\label{sec:training}

\subsection{Methods}

	The reconstructions presented in this work were trained using the three-layer scheme detailed above on experimental datasets of $N \approx 3000$ samples each. Training was performed using stochastic gradient descent with a decayed learning rate, the gradients being estimated via contrastive divergence with $k=30$ sampling steps. Since the visible layers of our machines are relatively small, exact computation of the negative-log-likelihood was possible on each set. Hyperparameters for training were therefore selected by cross-validation on a randomly chosen experimental set; the same hyperparameters were used in training on all datasets. The reconstructions presented in the text were trained on the full datasets; RBMs were also trained on 90/10 splits of each dataset in order to verify that the out-of-sample negative log likelihood did not grow during training. Error bars on reconstructed observables were computed from their variation across these training subsets in the final epochs of training. We found it beneficial to train each machine with the error rates set to zero for the first epoch.

\begin{figure}[h]
	\includegraphics[width=\columnwidth]{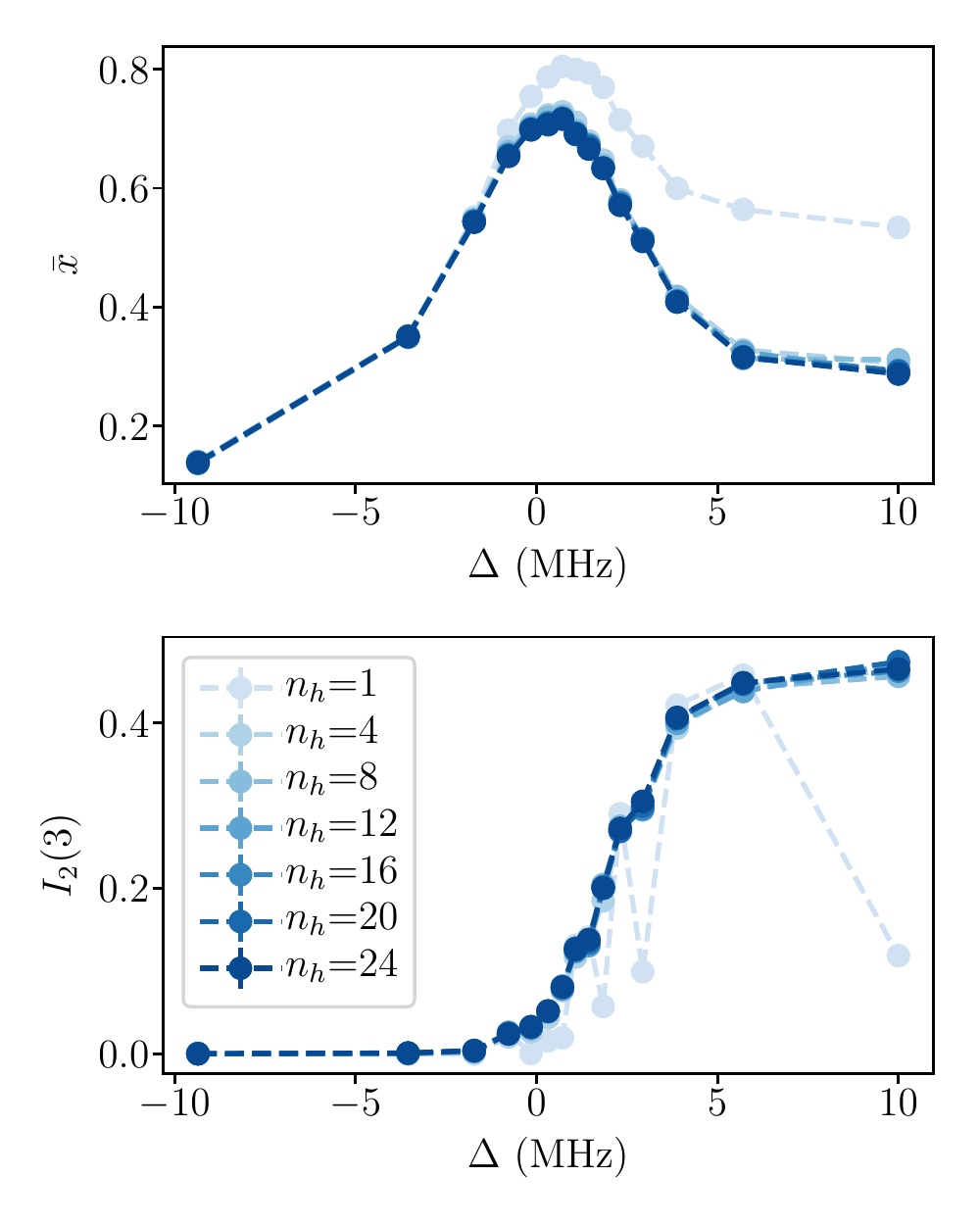}
	\caption{Examples of the scaling of observables with hidden layer size, for RBMs trained on experimental data. Top: spatially averaged transverse field values. Bottom: the Renyi mutual information at bond $s=3$. Error bars are defined by variation of reconstructed observables in the final epochs of training.}
	\label{fig:supp:nh_scaling}
\end{figure}

To check that the networks learned a consistent representation of the experimental data, we performed a scaling analysis of the number of hidden units $N_h$ of the RBM when training on experimental data. Increasing the number of hidden units, we found convergence of the observables and log-likelihood for $N_h \sim N$ (see Fig.~\ref{fig:supp:nh_scaling} for examples). The reconstructions presented in this work used RBMs with $N_h=2 N = 16$.

\subsection{Training on larger systems}
    As a test of the robustness of our reconstruction procedure, we also trained RBMs on a second set of Rydberg atom data sampled from a larger chain of $N=9$ atoms. The dynamics of this system is governed by a master equation identical in structure to that used for modeling the eight-atom data presented in the main text, but with slightly different detuning and Rabi frequency profiles, and different effective decoherence rates.
	
	\begin{figure}[h]
		\includegraphics[width=\columnwidth]{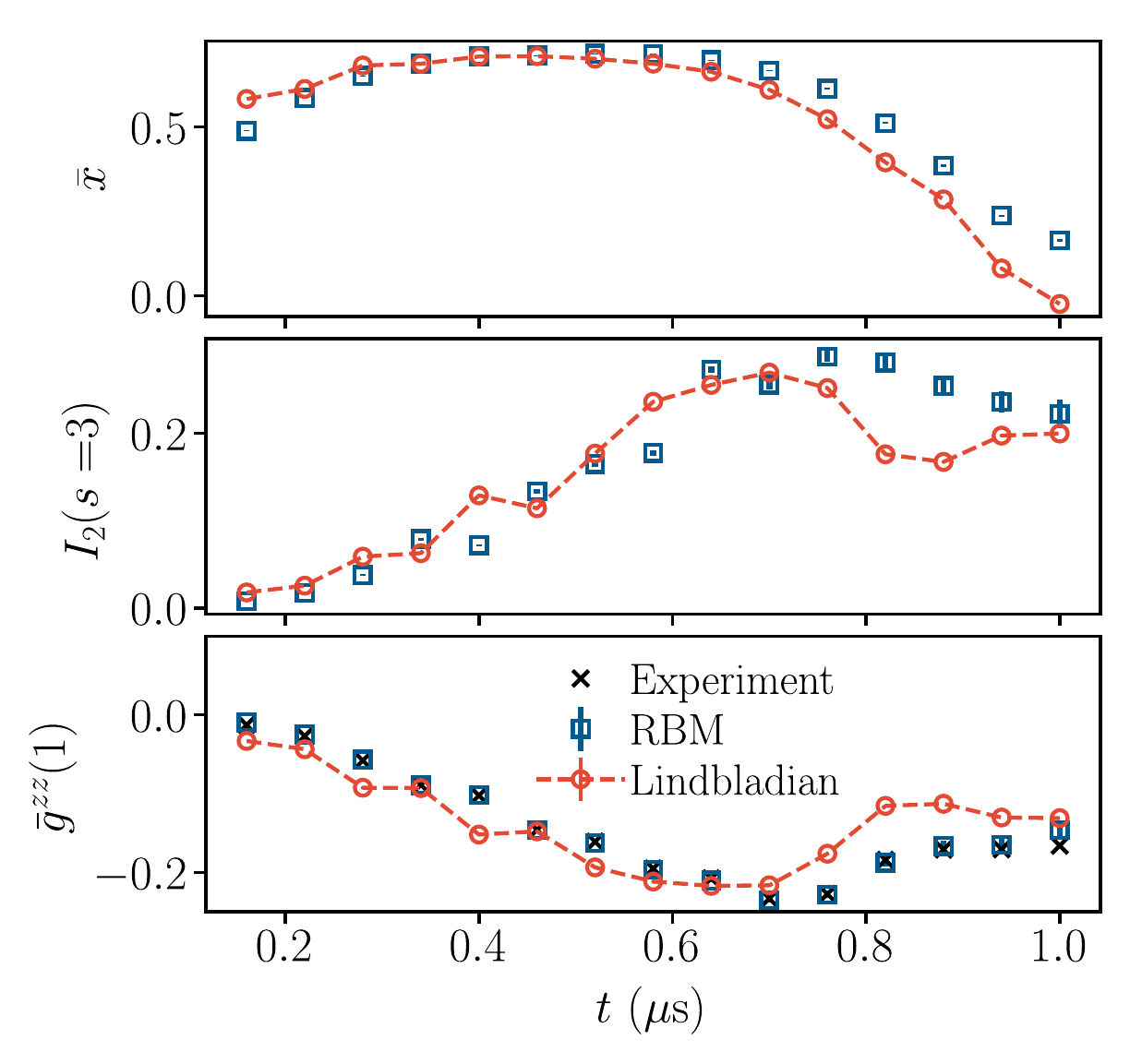}
		\caption{Some examples of observables reconstructed from nine-atom data, plotted as a function of sweep time $t$. From top to bottom: average transverse field $\bar{x}$, Renyi mutual information $I_2$ corresponding to a partition at bond $s=3$; averaged nearest-neighbor correlations in the measurement basis (including same noise model as in the main text). The machines were trained with the same hyperparameters as in the eight-atom case, using $N_h = 2N = 18$ hidden units.}
		\label{fig:nineatom-obs}
	\end{figure}

Fig.~\ref{fig:nineatom-obs} compares the results of this reconstruction to predictions of the relevant Lindbladian model, as well as experimental values where appropriate. Without alteration of the training procedure, the RBMs reconstruct quantum dynamics, as manifested in the transverse field and mutual information, in good agreement with Lindbladian predictions. This is a key benefit conferred to the experimentalist by the RBM reconstruction method. Indeed, given previous knowledge regarding the properties of the quantum state prepared in the experiment, RBM reconstruction of experimentally inaccessible observables allows for rapid and inexpensive detection of errors in state preparation and manipulation.

\section{Generalization capabilities}

A generative model is of little use if it merely mimics the statistics of the training set. Successful machine learning applications are built upon the ability to {\it generalize} from a given dataset, extracting representations of the data that capture relevant features of the ground truth distribution from which it was sampled. This requires some structure in the data for the machine to learn, and the extent to which it succeeds in doing so depends upon the architecture of the machine as well as the size of the dataset.

For relatively small datasets such as those used in this work, it is natural to wonder whether the apparatus of machine learning is necessary at all. In particular, given access to the frequency distribution (FD) 
\begin{equation}
	\pfd(\btau) = \frac{1}{N_s} \sum_{\btau_i \in \mathcal{D}} \delta_{\btau, \btau_i}
	\end{equation}
	defined by a particular dataset $\mathcal{D}$ consisting of $N_s$ samples, one may define a naive {\it frequency distribution reconstruction} of a pure state corresponding to the data, which simply memorizes the training set:
	\begin{equation}
	\ket{\Psi} = \sum_{\btau} \sqrt{\pfd(\btau)} \ket{\btau}
	\end{equation}

The FD state model can be computed and stored in a time linear in the size of the dataset, by building a lookup table that associates each observed bitstring $\btau$ with its empirical probability in the dataset, and assigning probability zero to all other bitstrings. Such a model may then be used to produce Monte-Carlo estimates of desired observables, in the same fashion as for RBM states.

In general, the FD reconstruction approach cannot scale to high-entropy distributions -- if $H_2$ is the second-order Renyi entropy of the ground-truth distribution $\pgt(\btau)$, the fidelity $F\left( \pfd, \pgt \right) = \sum_{\btau} \sqrt{\pfd(\btau) \pgt(\btau)}$  between the frequency distribution and the ground truth obeys the inequality

\begin{equation}
F\left( \pfd, \pgt \right) \le \sqrt{N_s} e^{-H_2/4}
\label{eqn:fidelity-bound}
\end{equation}
-- for a proof, see Section (\ref{section:appendix1}).
In particular, if the measurement-basis entropy is proportional to the system size -- as is the case in even some very simple states, such as a  
product state of spins not aligned with the measurement basis -- the frequency-distribution fidelity will decay exponentially in system size.
The ability to extract a modest number of {\it physically relevant features} is therefore essential for accurate state reconstruction from generic datasets of realistic size. However, our eight-atom system is small enough compared to the size of the datasets ($\sqrt{N_s} \sim 2^L$) that the FD approach is not {\it a priori} infeasible.

\begin{figure}[t]
\includegraphics[width=\columnwidth]{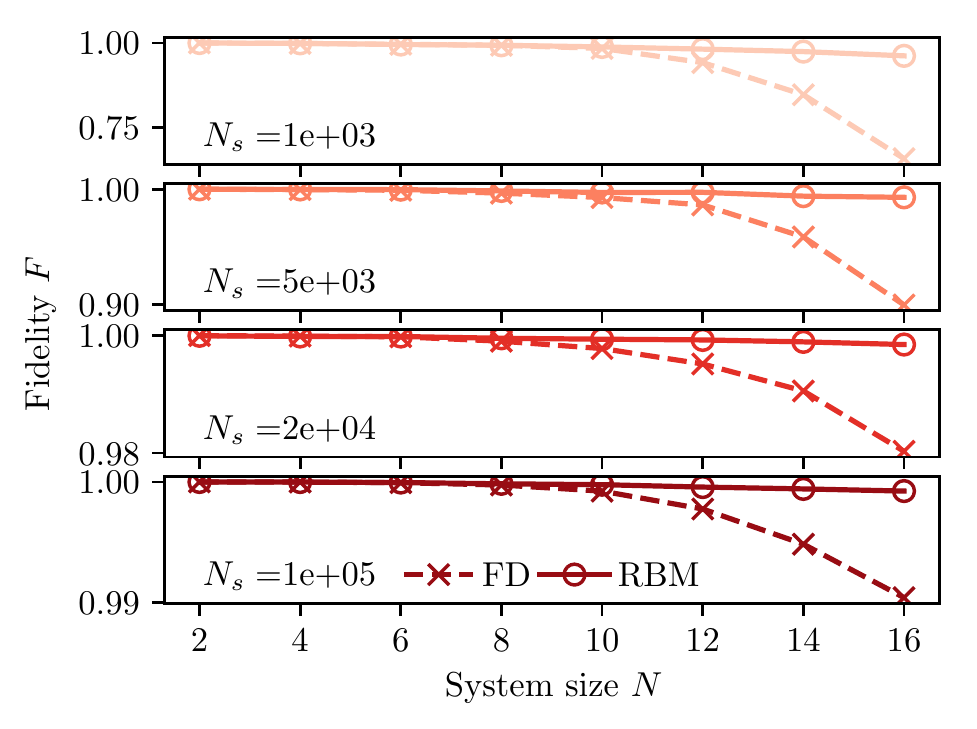}
\caption{Generalization from ground-state datasets: fidelity improvements conferred by RBMs over frequency-distribution reconstructions, for a selection of dataset sizes $N_s$. Note the change in scale.}
\label{fig:rbm-vs-fd-fidelity}
\end{figure}

To quantify the performance of RBM and FD reconstructions in the small-system regime, we sampled synthetic datasets (in the occupation number basis) of size $N_s$ up to $10^5$ from ground states of the Rydberg Hamiltonian in equation (\ref{eqn:Hryd}), for a selection of system sizes up to $N=16$ atoms. Ground state wavefunctions were computed using the QuSpin exact diagonalization package~\cite{Weinberg2017}; the Hamiltonian parameters were constant throughout and chosen to place the system near the phase transition into $\mathbb{Z}_2$ state: $V_{nn} = 30$MHz, $\Omega=2$MHz, $\Delta \approx 1$MHz. For each dataset, we computed the fidelity $F_{\textnormal{FD}}$ of the frequency distribution state onto the ground-truth Rydberg wavefunction; an RBM with $N_h = N$ hidden units was then trained on the same dataset, and its fidelity $F_{\textnormal{RBM}}$ onto the true state was also recorded. The RBMs were all trained with the  hyperparameters described in section~\ref{sec:training}, but with $k=10$ contrastive divergence steps. Fig.~\ref{fig:rbm-vs-fd-fidelity} plots the resulting fidelities achieved by both reconstructions as a function of system size -- RBMs of fixed complexity achieve significantly higher fidelities for large systems, with small improvements even at $N=8$.

Another issue of practical relevance is model size: given a dataset $\mathcal{D}$ of a particular size $N_s$, how many parameters are required to store each trained model? For the RBM, the number of (real-valued) parameters required to specify the model completely is determined by the size of the bias vectors and weight matrix, $N \cdot N_h + N + N_h$, and therefore quadratic in the system size for a fixed model complexity $N_h / N$. For the FD model, the number of parameters is determined by the size of the lookup table, i.e. the number of unique samples present in the dataset, and therefore bounded above by the dataset size $N_s$.

\begin{figure}[t] 
\includegraphics[width=\columnwidth]{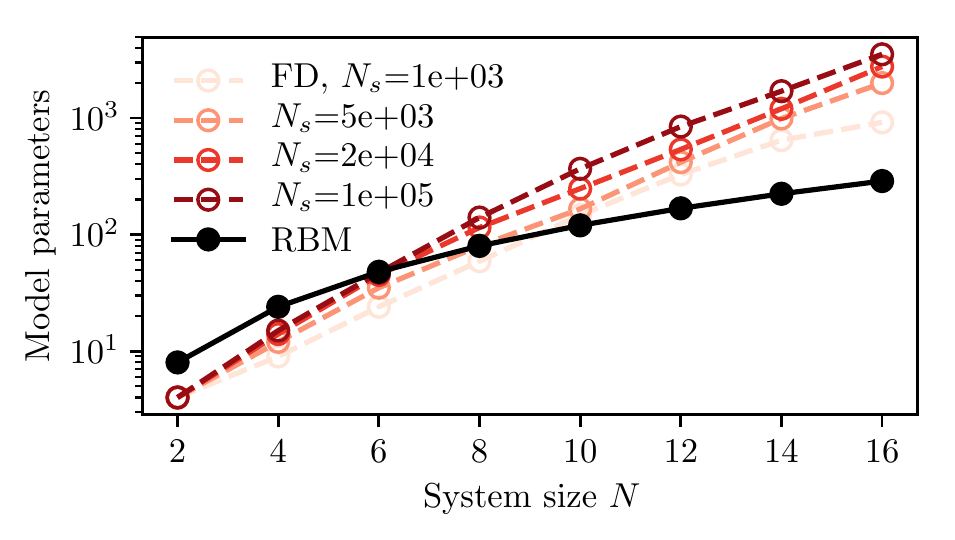}
\caption{Dependence of model size on physical system size (note the log scale). The solid line indicates the number of parameters
required to specify an RBM model with $N_h = N$. The dashed lines indicate the 
number of parameters required to build a lookup table for the FD model, for various dataset sizes $N_s$. }
\label{fig:param-count-scaling}
\end{figure}

In Fig.~\ref{fig:param-count-scaling}, the model sizes of the RBM and FD reconstructions from Fig.~\ref{fig:rbm-vs-fd-fidelity} are compared as a function of system size; for $N \gtrsim 8$ atoms the RBMs are a significantly more efficient (not to mention more accurate) description of the quantum state.

Finally, we note that even for small systems, generative models provide an additional advantage in state reconstruction from noisy data: in the presence of measurement errors, the FD model is not representative of the ground truth for any dataset size, and simply inverting the conditional probabilities will generally result in unphysical prior distributions. Denoising methods for cleaning noisy binary datasets prior to reconstruction~\cite{Greig1989,Geman1984,Batson2019} may be applied, but a model training step is still required. 

\section{Effects of Decoherence}
	For pure state reconstruction to be useful in near-term quantum simulators, realistic decoherence processes must be accounted for. Here, we provide a brief description of the Lindbladian master equation used in our modeling of the experiment, and discuss means of assessing the quality of pure state reconstructions in the presence of decoherence.

\subsection{Master equation for the Rydberg machine}

   To account for decoherence processes quantitatively, we have used a Lindblad model, described in detail in Ref.~\cite{Levine2018}, which includes two jump operators $\tilde \sigma_i^{rg} = \ket{g}\bra{r}, \tilde \sigma_i^{gg} = \ket{g}\bra{g}$ to represent decay and dephasing processes acting on atom $i$. The time evolution of the full state is given by the master equation
\begin{equation}
\begin{split}
\frac{d \hat \rho}{dt} &= -i [ \hat H(\Omega(t), \Delta(t))
+ \hat H_{dis}, \hat \rho]\\ &+ \sum_{i=1}^N \sum_{t=rg,gg} \gamma_t \left( \tilde \sigma_i^t \:\hat \rho\: \tilde \sigma_i^{t \dag} - \frac{1}{2} \left\{ \tilde \sigma_i^{t \dag} \tilde \sigma_i^{t}, \hat \rho  \right\} \right)
\label{eqn:ME}
\end{split}
\end{equation}
	 where $\hat H_{dis} = -\sum_{i=1}^N \delta_i \hat n_i$ is the static disorder Hamiltonian containing the Doppler shifts $\delta_i$, and $\gamma_t$, $t=rg, gg$ are decoherence rates estimated from single-atom measurements~\cite{Levine2018} as $1/\gamma_{rg} =$80$\mu$s, $1/\gamma_{gg}=$40$\mu$s respectively. The Doppler shifts $\delta_i$ were assumed to be Gaussian-distributed with an rms width of $2 \pi \cdot 43.5$kHz. Direct spontaneous decay processes from the Rydberg states, which occur over longer timescales, were neglected. Numerical solutions of the master equation (\ref{eqn:ME}) were performed using QuTiP~\cite{qutip}, and observables were averaged over 100 disorder realizations $\{ \delta_i \}$. Uncertainties in observables were computed from the standard error of the mean of these realizations. We note that the experiment has additional loss mechanisms, as well as imperfections in the laser sweep profile, which are not well characterized and not included in this Lindbladian model. We believe this accounts for the discrepancy with experimental correlation functions noted in the main text.
	 
	 \begin{figure}[t]
	 	\includegraphics[width=\columnwidth]{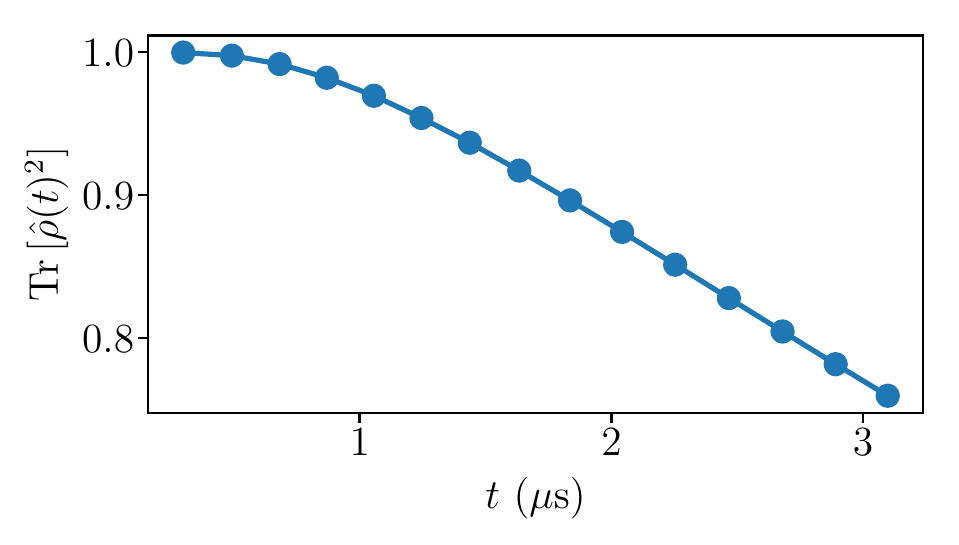}
	 	\caption{Purity of the master equation solutions as a function of sweep time $t$.}
	 	\label{fig:lindblad-purity}
	 \end{figure}

	 This master equation predicts a substantial loss in purity $\tr \left[ \hat \rho^2 \right]$ for states produced at the end of the sweep (Fig.~\ref{fig:lindblad-purity}), whose detrimental effects on our pure-state reconstruction process we quantify below.

\subsection{Reconstruction fidelities}
 To assess the quality of quantum state reconstruction, we consider the fidelity between two states $\hat \rho, \hat \sigma$,
	\begin{align}
	    F(\hat \rho, \hat \sigma) = \tr{ \left[ \sqrt{ \sqrt{\hat \rho} \hat \sigma \sqrt{\hat \rho} } \right]     } 
	\end{align}
	which reduces to the norm of the overlap in the case where $\hat \rho, \hat \sigma$ are pure states. An ideal state reconstruction $\hat \sigma$ of a mixed state $\hat \rho$ would yield $F(\hat \rho, \hat \sigma) =1$. For pure state reconstructions  $\hat \sigma = \ket{\psi_{\bm{\lambda}}} \bra{\psi_{\bm{\lambda}}}$, this is not possible if the true state $\hat \rho$ is non-pure. However, one may still seek an approximate reconstruction which reproduces the local reduced density operators of $\hat \rho$. In particular, specializing to the case of one-dimensional systems, we can consider contiguous subsystems formed from $s$ adjacent sites, $A^{(s)}_i = \{i, i+1, ..., i+s-1 \}$. Given two density operators $\hat \rho, \hat \sigma$ for the global system $\mathcal{S}$ of size $N$, the reduced density operators which describe the subsystem in each state are obtained by tracing out the rest of the chain, 

\begin{align*}
	\hat \rho^{(s)}_i &= \tr_{\mathcal{S} / A^{(s)}_i} \left[ \hat \rho \right] \\
	\hat \sigma^{(s)}_i &= \tr_{\mathcal{S} / A^{(s)}_i} \left[ \hat \sigma \right] 
\end{align*}
Then we define a \textit{subsystem averaged fidelity} as the spatial average of the fidelity between these local operators, over all subsystems of a particular size $s$:
	\begin{align}
		F_s \left(\hat \rho, \hat \sigma \right) = \frac{1}{N+s-1} \sum_{i=1}^{N-s+1} F \left(\hat \rho^{(s)}_i , \hat \sigma^{(s)}_i \right)
		\end{align}
$F_s \left(\hat \rho, \hat \sigma \right)$ is a measure of how well, on average, $\hat \sigma$ is able to reproduce the $s$-local physics of $\hat \rho$.

	To examine the quality of the RBM states $\hat \rho_{\bm{\lambda}} = \ket{\psi_{\bm{\lambda}}} \bra{\psi_{\bm{\lambda}}}$ in reproducing local density operators, we solved the master equation (\ref{eqn:ME}) for set of decay rates $\gamma_{rg} = \alpha \gamma_{rg}^{\textnormal{exp}}, \gamma_{gg} = \alpha \gamma_{gg}^{\textnormal{exp}}$, with $\gamma_{rg}^{\textnormal{exp}}, \gamma_{gg}^{\textnormal{exp}}$ denoting our estimates of the experimental values, and $\alpha$ a dimensionless parameterization of the overall decoherence strength. For each set of decoherence rates, the master equation was solved and synthetic data sampled from the resulting mixed states. Pure state RBMs were trained on each of these datasets, and the resulting averaged fidelities $ F_s \left(\hat \rho, \hat \rho_{\bm{\lambda}} \right)$ were computed. 

\begin{figure}[t]
	\includegraphics[width=\columnwidth]{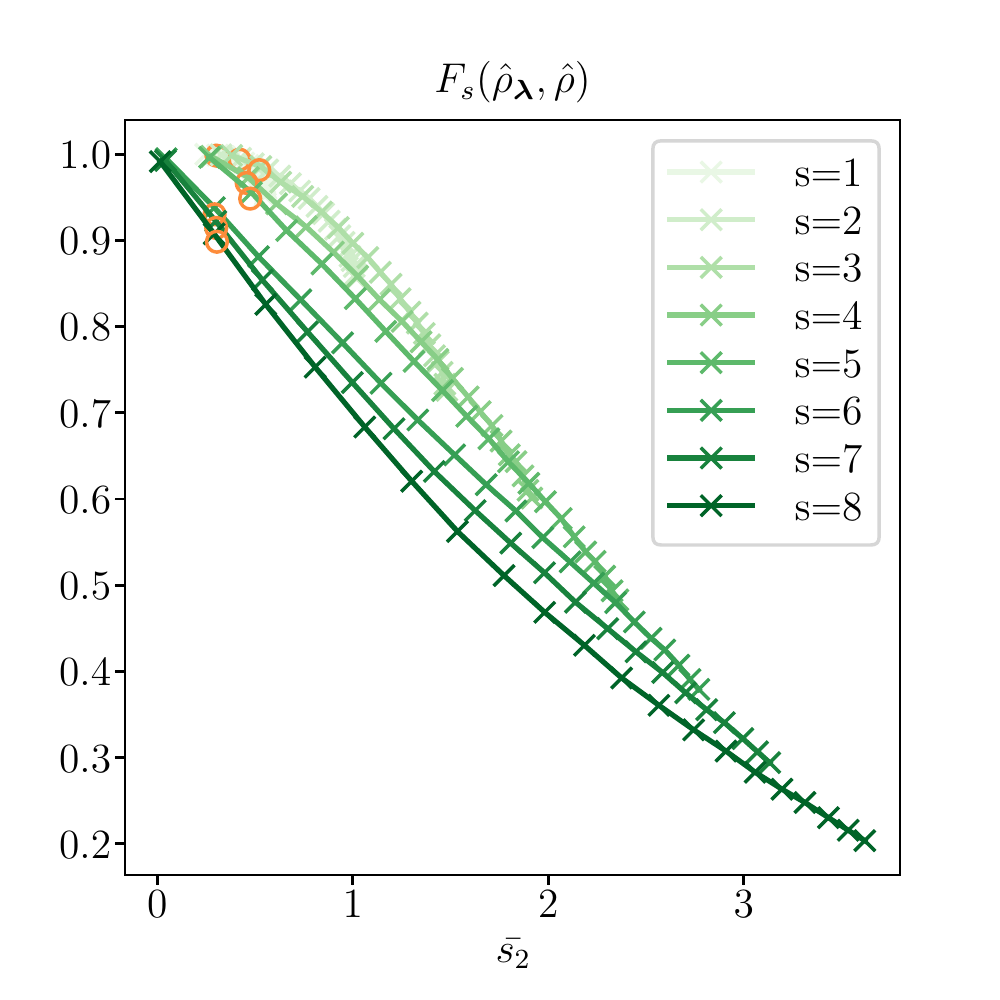}
	\caption{\textbf{Average subsystem fidelities.} For each subystem size $s$, the average subsystem fidelity between the reconstructed state $\hat \rho_{\bm \lambda}$ and the state $\hat \rho$ from which its training data was sampled is plotted, for varying values of the decoherence rates, as quantified by the average Renyi entropy $\bar{s_2} = - \frac{1}{N+s-1} \sum_{i=1}^{N-s+1} \log \tr\left( \hat \rho_i^{(s) 2}\right)$ of the local reduced density operators.
 The data plotted are for states taken at the end of the sweep, at $\Delta = 10$MHz. Open circles indicate the fidelities obtained using the decoherence rates from the experimental model presented in the main text. The fidelity behavior at other points in the sweep (not shown) is qualitatively similar.}
	\label{fig:supp:fidelity_vs_entropy}
\end{figure}
As a representative example, Fig. \ref{fig:supp:fidelity_vs_entropy} shows how the fidelities computed in the final state of the sweep vary as a function of the average Renyi entropy $\bar{s_2}$ of subsystems of a given size -- one observes a roughly linear decay in the average fidelity with the averaged entropy. These numerical results suggest that pure state reconstruction techniques should focus on few-body operators, where the entropy build-up due to global decoherence process is limited in proportion to the system size. 

\section{Reconstruction improvement from noise layer regularization}

	Numerical experiments have demonstrated that noise layer regularization results in higher-fidelity pure state reconstruction when training on uniformly noisy data.
	
	\begin{figure}[t]
		\includegraphics[width=\columnwidth]{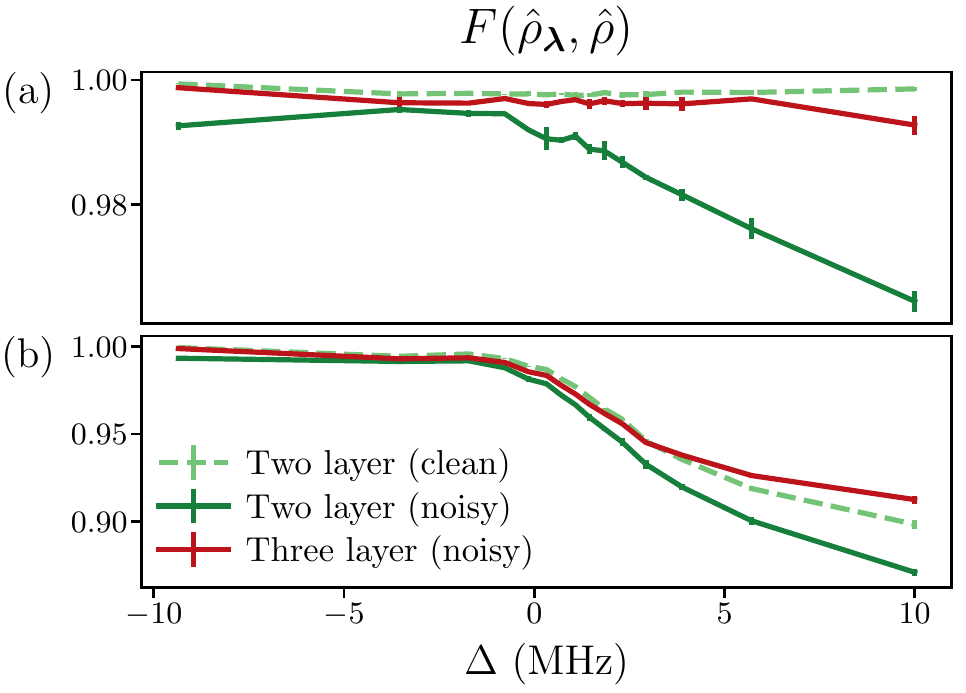}
		\caption{\textbf{Fidelity improvements from noise layer regularization}. As a demonstration of the efficacy of noise layer regularization, we plot the fidelity $F(\hat \rho_{\bm \lambda}, \hat \rho)$ obtained between the underlying state $\hat \rho$ and the reconstruction $\hat \rho_{\blda} = \ket{\psi_{\blda}} \bra{\psi_{\blda}}$, when training on synthetic data subjected to measurement errors (`noisy' data), as a function of detuning $\Delta$. We compare regularized training (red solid lines, `Three Layer') with unregularized training (green solid lines, `Two Layer'), for (a) Data sampled from pure, positive Rydberg ground states, and (b) Data sampled from the mixed states $\hat \rho$ predicted by our Lindbladian model. As a benchmark we plot in each case the fidelity obtained by a two-layer RBM training on `clean' data without measurement errors (green dashed lines). The regularized training leads to higher fidelities for all states sampled. For some mixed states, it even exceeds the RBM trained on clean data. This is because the pure state model is no longer valid when the source state is mixed, and so the `optimal' pure state as defined by fidelity is not necessarily the one which best fits the training set. }
		\label{fig:noise-layer-regularization-fidelities}
		\end{figure}

	 Fig.~\ref{fig:noise-layer-regularization-fidelities} compares fidelities achieved by regularized and unregularized RBMs, when trained on synthetic datasets subjected to the bitflip error channel described in the main text.  The improvement is significant, especially in the ordered phase, where global state purity is lowest. It is important to note that in these experiments the noise process is known ahead of time and built into the three layer networks as in Fig.~\ref{fig:supp:threelayer_schematic}. We have also trained three-layer machines using incorrect values of the error rates on the same noisy synthetic data. Although the fidelity performance varied somewhat, depending most sensitively on $p(0|1)$, the quality of the regularized reconstructions is generally robust, and deep in the ordered phase all three-layer machines exhibited higher fidelities than their two layer counterparts on the corresponding datasets, for error rates with bounds set by single-atom measurements~\citep{Levine2018}. Generically, of course, a sufficiently large mismatch between the true and assumed error rates will lead to decreased reconstruction fidelity. Future work will investigate more generally the task of selecting a regularization method for noisy quantum data.
			\begin{figure}[t]
			\includegraphics[width=\columnwidth]{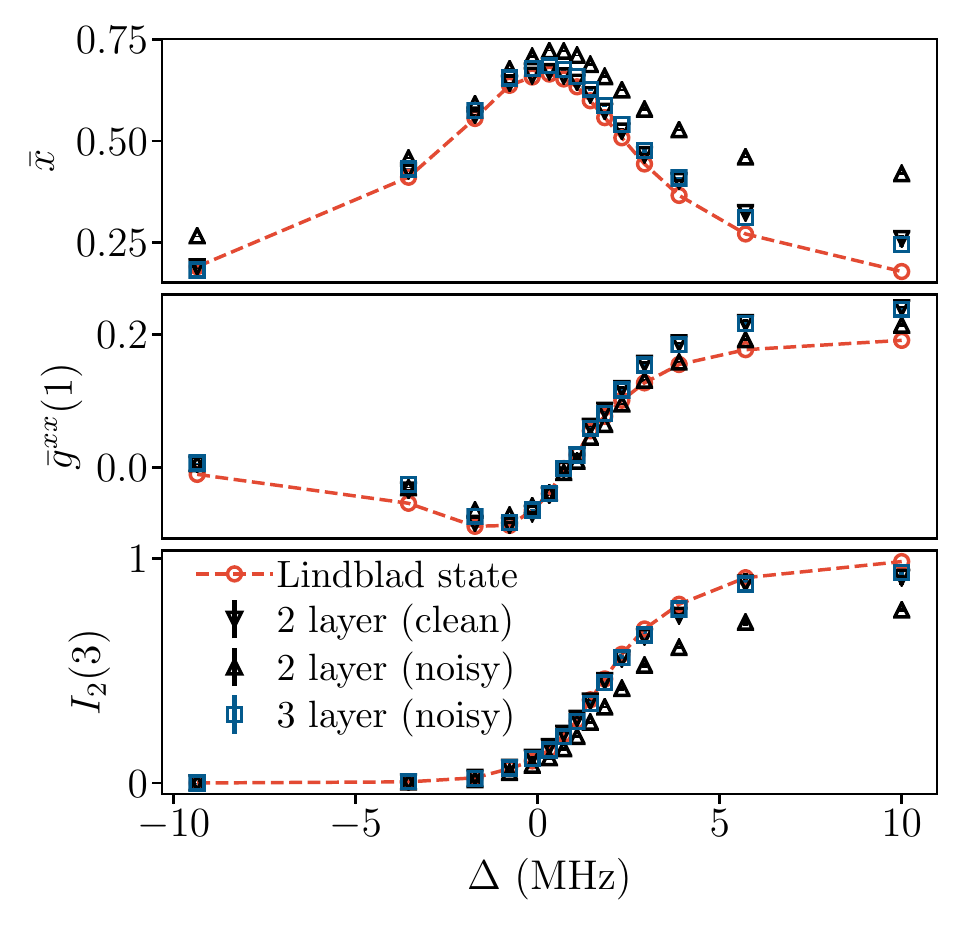}
			\caption{\textbf{Observable reconstructions from synthetic data}. A comparison of two- and three-layer reconstructions of the Lindbladian state when subjected to measurement errors. From top to bottom: average transverse field, average nearest-neighbor $XX$ correlation, and Renyi mutual information at bond 3. `noisy' (`clean') indicates training data with (without) measurement errors. Note the close agreement between the three layer machines trained on noisy data (blue squares) and the two-layer machines trained on clean data .}
		\label{fig:three_layer_obs}
		\end{figure}

 Fig.~\ref{fig:three_layer_obs} compares the predictions of these synthetically trained two- and three-layer machines (using the known noise values) for some of the observables discussed in the main text. We find that noise-layer training allows the RBMs to provide much tighter agreement in, for example, values of the transverse field and mutual information. Surprisingly, the three-layer machines actually produce poorer estimates of the transverse field correlator in the ordered phase, despite yielding two-body density operators with higher fidelities for all sampled states. A more detailed analysis of the ordered phase states reveals that regularized training does indeed produce better estimates of the one- and two-body expectation values $\langle \sigma_i^x \rangle$, $\langle \hat \sigma_i^x  \hat \sigma_{i+1}^x \rangle$ for all sites $i$. However, at bonds (3, 4) and (5, 6), where quantum fluctuations are strongest, the three-layer improvement in the two-body expectation value is relatively small, while the reduction in the one-body expectation value is substantial. Upon computing the connected correlator $\langle \hat \sigma_i^x \hat \sigma_{i+1}^x \rangle - \langle \sigma_i^x \rangle \langle \sigma_{i+1}^x \rangle$, the overall effect is an overestimate of the true correlation.

\section{Appendix: Proof of classical fidelity bound}
\label{section:appendix1}

Inequality (\ref{eqn:fidelity-bound}) is obtained by bounding the probability of the most likely outcome using the Renyi entropy $H_2$ in the measurement basis. By definition, $H_2 = - \log \sum_{\btau} \pgt(\btau)^2$, and $ \sum_{\btau} \pgt(\btau)^2 \ge \max_{\btau} \pgt(\btau)^2$, so $-\log \sum_{\btau} \pgt(\btau) ^2 \le -2 \log \max_{\btau} \pgt(\btau)$. Rearranging, $\max_{\btau} \pgt(\btau) \le e^{-H_2 / 2}$. In particular, this bounds the probability of any event in the training set, so 
\begin{align*}
  F(\pfd, \pgt) &= \sum_{\btau} \sqrt{ \pfd(\btau) \pgt(\btau) } \\
  					&\le \sum_{\btau} \sqrt{ \pfd(\btau) e^{-H_2/2}} \\
  					&= \sqrt{N_s} e^{-H_2/4}
\end{align*}

\section{Appendix: Renyi entropy bound from positive pure states}

\newcommand{\rhotrue}{\hat \rho}
\newcommand{\rhopp}{\hat \rho^P}
\newcommand{\rhoA}{\rhotrue_A}
\newcommand{\rhoppA}{\rhopp_A}
\newcommand{\wf}[3]{\Psi_{#1,#2}^{#3}}

The $n$th order Renyi entropy of a quantum state $\rhotrue$ is defined as $S_n \left[ \rhotrue \right] = \frac{1}{1-n} \log \tr \rhotrue^n$.

Consider a system $S$ partitioned into subsets $A$ and $B$, and
a density operator $\rhotrue$ defined on $S$; its reduced
density operator in the $A$ subsystem is $\rhoA =\textrm{Tr}_{B}\rhotrue$. Let $\ket i$, $\ket j$ denote orthonormal bases for $A,B$ respectively, so that the set of product states $\ket{i,j}$ forms an orthonormal basis for the full system $S$. Let $p_{i,j}$ be the probability assigned by $\rhotrue$ to the measurement outcome $i,j$: $p_{i,j} = \tr \left( \rhotrue \ket{i,j} \bra{i,j} \right)$. The positive-pure partner to the mixed state is defined as 

\begin{align}
\ket{\Psi^P[\rho]}=\sum_{i,j} \sqrt{p_{i,j}}\ket{i, j}, 
\end{align}

and the corresponding reduced density operator on $A$ is $\rhoppA = \textrm{Tr}_{B} \ket{\Psi^P[\rho]} \bra{\Psi^P[\rho] }$.

{\bf Theorem: } For $n > 1$, the Renyi entropies $S_n$ of the two density operators satisfy the inequality
\begin{align}
S_n \left[  \rhoppA \right] \le S_n \left[  \rhoA \right]
\end{align}

As a consequence, in the case of pure states $\rhotrue$, where the global Renyi entropy vanishes, the positive-pure partner provides a lower bound on the mutual information:
\begin{align}
 	I_n \left[ \rhopp \right] &= S_n \left[ \rhopp_A \right] +  S_n \left[ \rhopp_B \right] \\
 								      &\le S_n \left[ \rhotrue_A \right] +  S_n \left[ \rhotrue_B \right]  \\
 								      &= I_n \left[ \rhotrue \right]
 \end{align}

We note that for the case of the $n=2$ Renyi entropy and pure states $\rhotrue$, this result has been obtained in previous work~\cite{Zhang2011,Grover2015}.

{\bf Proof:} Choose an auxiliary system $R$ to purify $\rhotrue$: $\rhotrue=\textnormal{Tr}_{R}\ket{\Psi}\bra{\Psi}$ for some pure state
$\ket{\Psi}$ living in $S\otimes R$. If $\ket{\alpha}$ is an orthonormal basis for $R$, we can expand the 
larger pure state in the joint basis $\ket{i, j,\alpha}$ : $\ket{\Psi}=\sum_{i,j,\alpha} \wf{i}{j}{\alpha} \ket{i,j,\alpha}$ for some complex coefficients $\wf{i}{j}{\alpha}$.

In terms of these amplitudes, the reduced density operator of the mixed state on $A$ is
\begin{equation}
\rhoA =\sum_{\alpha,j} \wf{i}{j}{\alpha} \wf{i'}{j}{\alpha *} \ket{i}\bra{i'}
\end{equation}

and so 
\begin{equation}
\tr{\rhoA^{n}}=\left( \wf{i_1}{j_1}{\alpha_1} \wf{i_2}{j_1}{\alpha_1 *} \right) \left( \wf{i_2}{j_2}{\alpha_2 } \wf{i_3}{j_2}{\alpha_2 *} \right) \dots \left( \wf{i_n}{j_n}{\alpha_n} \wf{i_1}{j_n}{\alpha_n *} \right)
\end{equation}

with summation over all indices implied. The reduced density operator for positive-pure partner may be obtained from the definition above:
\begin{equation}
\rhoppA =\sum_{i,i',j}\sqrt{p_{i,j}p_{i',j} }\ket{i}\bra{i'}
\end{equation}
whence 
\begin{equation}
\tr \left( \rhoppA \right)^n = \left(\sqrt{p_{i_1,j_1}p_{i_2,j_1} } \right)  \left( \sqrt{p_{i_2,j_2}p_{i_3,j_2} } \right) \dots \\
 \left( \sqrt{p_{i_n,j_n}p_{i_1,j_n} } \right)
\end{equation}

(summation implied). Furthermore, 
\begin{align}
p_{i,j} & =\sum_{\alpha} \wf{i}{j}{\alpha} \wf{i}{j}{\alpha *} 
\end{align}
and so by the Cauchy-Schwartz inequality,
\begin{align}
   \left| \sum_\alpha \wf{i}{j}{\alpha} \wf{i'}{j'}{\alpha*} \right| &\le \sqrt{  \left( \sum_\alpha \wf{i}{j}{\alpha} \wf{i}{j}{\alpha*} \right)
    \left( \sum_{\alpha'} \wf{i'}{j'}{\alpha'} \wf{i'}{j'}{\alpha'*} \right) } \\
    &= \sqrt{p_{i,j} p_{i',j'} }
    \end{align}
   
Therefore, writing $\mathbf{i} = (i_1, ..., i_n), \mathbf{j} = (j_1, ..., j_n)$, 
\begin{align}
\tr{\rhoA^{n}} &=\sum_{\mathbf{i}, \mathbf{j}} \left(\sum_{\alpha_1} \wf{i_1}{j_1}{\alpha_1} \wf{i_2}{j_1}{\alpha_1 *} \right) \left( \sum_{\alpha_2} \wf{i_2}{j_2}{\alpha_2 } \wf{i_3}{j_2}{\alpha_2 *} \right) \nonumber \\ 
          & \dots \left( \sum_{\alpha_n} \wf{i_n}{j_n}{\alpha_n} \wf{i_1}{j_n}{\alpha_n *} \right) \\
          &\le \sum_{\mathbf{i}, \mathbf{j}} \left| \sum_{\alpha_1} \wf{i_1}{j_1}{\alpha_1} \wf{i_2}{j_1}{\alpha_1 *} \right| \left| \sum_{\alpha_2} \wf{i_2}{j_2}{\alpha_2 } \wf{i_3}{j_2}{\alpha_2 *} \right| \nonumber \\
          & \dots \left| \sum_{\alpha_n} \wf{i_n}{j_n}{\alpha_n} \wf{i_1}{j_n}{\alpha_n *}  \right|  \\
          &\le  \sum_{\mathbf{i}, \mathbf{j}}  \left(\sqrt{p_{i_1,j_1}p_{i_2,j_1} } \right)  \left( \sqrt{p_{i_2,j_2}p_{i_3,j_2} } \right) \dots 
 \left( \sqrt{p_{i_n,j_n}p_{i_1,j_n} } \right) \\
 	    &= \tr \left( \rhoppA \right)^n
				\end{align}
	 Hence $-\log \tr \left( \rhoppA \right)^n \le -\log  \tr \left( \rhoA \right)^n$, which means that for $n>1$, $S_{n}(\rhoppA )\le S_{n}(\rhoA)$. 

\end{document}